\documentclass[12pt]{article}
\usepackage[utf8]{inputenc}
\usepackage[T1]{fontenc}
\usepackage{mathtools}
\usepackage{ragged2e}
\usepackage{color,soul}
\usepackage{tabularx}
\usepackage{eurosym}
\usepackage[english]{babel}
\usepackage[margin=1in]{geometry}
\usepackage{amsmath,amsthm,amssymb}
\usepackage{indentfirst}
\usepackage{hyperref}
\usepackage{ulem}
\usepackage{setspace}
\usepackage{adjustbox}
\usepackage{rotating}       
\usepackage{threeparttable} 
\usepackage{subcaption}         
\usepackage{booktabs}       
\usepackage{array}
\usepackage{float} 
\usepackage[]{caption}
\usepackage[justification=justified,singlelinecheck=false, font={small},labelfont={bf}]{caption}
\usepackage{booktabs}    
\usepackage{siunitx}     
\usepackage{subcaption}

\newcommand{\setfoot}[2]{%
    \footnote{#2}%
    \newcounter{#1}%
    \setcounter{#1}{\value{footnote}}%
}

\newcommand{\qedwhite}{\hfill \ensuremath{\Box}}

\newcommand\numeq[1]%
  {\stackrel{\scriptscriptstyle(\mkern-1.5mu#1\mkern-1.5mu)}{=}}
  
\title{Consumer Choice over Shopping Baskets}
\author{Afonso Rodrigues\setfoot{unia}{Rodrigues: Department of Economics, University of Oxford (afonso.rodrigues@economics.ox.ac.uk). I have no relevant or material financial interests that relate to the research described in this paper. I gratefully acknowledge financial support by the Foundation for Science and Technology of the Ministry of Education, Science and Innovation (Portugal), as well as data sharing by SONAE. The data agreement involves a request for review of the findings prior to their release. I am indebted to my dissertation advisor Alexei Parakhonyak, as well as Ian Crawford, Alex Teytelboym, Howard Smith, and Giulio Gottardo for their helpful insights. I benefited from thoughtful conference discussions by Leonardo Madio, Anne-Christine Barthel, and Alminas Zaldokas, as well as valuable comments from participants of the NIE PhD Symposium 2024, the 13th Oligo Workshop, the Research Jamboree at the University of Oxford, the Lisbon Meetings in Game Theory and Applications \#14, EARIE 2025, the CEPR Paris Symposium 2025, MaCCI 2026, and the Empirical IO Workshop at the University of Bristol.}}

\date{\today}

\begin{document}
\maketitle
\small

\begin{center}
    \textbf{Abstract}\\
\end{center}

I introduce a novel approach to structural modelling and estimation of continuous demand systems, utilising \textit{consideration sets} to analyse differentiated products markets with very large choice sets and purchases over multiple goods, multiple units, and across product categories. I apply it to study intra-store competition in the Portuguese supermarket industry between 2020 and 2023, during which the country faced the COVID pandemic. Anonymised transaction-level point-of-sale data is sufficient to estimate price elasticities across almost 30 000 goods and more than 500 product categories. Results suggest mark-ups remained stable throughout the sample period, with a short-lived, slight increase post-pandemic observed only in the highest-mark-up-percentile goods. The implied mark-ups match observed price volatility, profit margin surveys, as well as reports on shifting consumer tastes during the sample period.\\



\normalsize
\pagebreak
\section{Introduction}

Demand estimation provides a picture of how buyers react to prices and product features. It allows evaluations of market power, merger assessments, and the tracking of spillovers, as well as gauging the effects of taxes, subsidies, or new rules before they are put in place. The research goal of this paper is to apply demand estimation to study intra-store competition in the Portuguese supermarket industry. For this purpose, I estimate a structural demand model to obtain price elasticities and analyse mark-up trends across product categories. I obtained point-of-sale data from across a sample of grocery stores owned by a supermarket chain in Portugal, for the period between February 2020 to February 2023. This setting and time period make the analysis especially relevant for understanding price formation and cost pass-through in retail markets during a period of significant socio-economic disruption. The dataset covers almost 30000 goods and more than 500 product categories but is anonymised, requiring the estimation of a model of aggregate demand. \\


The complex setting of purchases in supermarkets - multi-unit, multi-product, cross-category - favours the use of a continuous demand system. Continuous demand allows choice over consumption bundles of many goods, and can incorporate both complementary and substitute alternatives. I work throughout with a quasilinear quadratic utility specification, as a general second-order local approximation to any well-behaved quasilinear utility function, which yields linear demand.\\

Aggregate linear demand raises several problems for structural estimation, which have prevented its wider use in empirical research. Two of these are common to all continuous demand systems (Gandhi and Nevo, 2021). First is a curse of dimensionality; as the number of alternatives grows larger, the number of own- and cross-price effect parameters grows (proportional to) quadratically in these. For large enough product assortments, these may outnumber the data points, preventing identification. Second is price endogeneity, which arises from the simultaneity of demand and supply functions. For each product, linear demand incorporates many prices beyond a product's own. Overcoming price endogeneity concerns requires instrumenting each price individually using many exogenous and non-collinear instruments to separately identify price effects.\\

A further well-known issue specific to linear demand is that aggregation of individual linear demand functions fails to discipline structural parameters. It does not hold generally that there exists a representative consumer whose utility-maximising demand for goods rationalises the observed aggregate demand system. This is due to consumer heterogeneity; consumers with different preferences have different reservation prices. Aggregation combines demands from these different regimes. The resulting aggregate demand - which for a given price may be at a corner, zero-demand for a consumer yet not for another - need not be globally rationalised as that of a representative consumer. Without a representative consumer, it is not possible to pin down how the set of active consumers per good is affected by changes in prices, so structural parameters are not identified. These corner solutions are a function of the prices and therefore cannot be instrumented away. They introduce a form of price endogeneity: as prices rise, the necessary adjustment to force negative demands back to zero also rises.\\

Another issue arises which is more context-dependent: consumer consideration of every feasible consumption bundle in the supermarket is unfeasible. For example, guacamole is a popular spread which most commonly mixes avocado, red onion, and lime. These are items which are otherwise rarely bought in pairs; consumers tend to only consider either the combination or the individual goods. The existence of consumer-specific consideration sets over combinations of goods, which differ from the set of feasible alternatives, is a form of rational inattention. It challenges the idea of preferences and choices over goods. The decision to buy the combination of avocados, red onions, and limes implies demand for each good is dependent on that of the others via co-consideration, beyond substitution or complementarity links. This dependence may have implications for the outcomes of consumer choice, and must therefore be modelled directly. \\

In this paper, I propose a theory of consumer choice, whereby consumers' individual consumption bundles are constrained by the choices they make over finite, latent consideration sets. These consideration sets are made up of different unique combinations of goods, which I call \textit{shopping baskets}. Consumers maximise utility through their choice of shopping baskets, meaning their effective consumption bundle is constrained to the vector space enabled by the consideration set. This accommodates broad flexibility in consumer choice while imposing realistic constraints. My empirical approach then applies this theoretical model to study intra-supermarket competition, in the process addressing both the technical and computational challenges of estimating aggregate linear demand.\\

The first of this paper's key contributions is an aggregation result. I formalise when aggregate demand from consumers with heterogeneous preferences and consideration sets can be rationalised by a representative consumer. The aggregate demand is piecewise linear, and the representative consumer solves a constrained maximisation problem within the same local functional class. This is valuable as it means that the model remains empirically and theoretically useful even when only market-level data is observed; it provides conditions under which underlying population preference parameters can be learned via structural estimation. Gorman (1953, 1961) admits consumer heterogeneity only if consumers do not choose corner solutions. My proposed result enables deep heterogeneity in preferences, price sensitivity, consideration sets, and reservation prices. The only conditions for the result to hold are that (i) local demand slopes be proportional across all consumers, and (ii) a sufficiently rich aggregate consideration set be observed, obtained from a combination of unique shopping baskets across all consumers' consideration sets. The latter is later shown to hold in general for even just a small number of consumers, while the former corresponds to an assumption that there be an approximate agreement amongst consumers as to which goods are or not substitutes. This contribution turns out to be informative for addressing price endogeneity due to corner solutions, enabling the separation of price effects from changes in the active set of consumers.\\

The second key contribution is the enabling of demand estimation via a model of linear demand, addressing of the Gandhi and Nevo (2021) critiques. To handle the curse of dimensionality, I expand on the Pinkse and Slade (2004) semi-parametric demand estimation approach, proxying for substitution effects via a statistical model of joint-purchase frequencies due to Tian, Lautz, Wallis, and Lambiotte (2021). The statistical model adjusts for product popularity and transaction size and is fed into the demand model via polynomial series expansion to predict price effects across products. It allows flexibility to observe both substitutes and complements. The empirical model is Two-Stage Least Squares, with a set of linear demand functions, one per good with supermarket and time fixed effects (FEs). Product utility is a function of observed product characteristics, including product category and private-label brand dummy variables. Prices are interacted with these to allow for heterogeneity. The proxy polynomials are interacted with a single exogenous instrumental variable - the average price across each goods' competitors at other grocery stores (Hausman, 1996). This covers the price endogeneity concerns from the simultaneity of demand and supply.\\

The third key contribution is the empirical application itself. Using the model above, I am able to recover mark-up measurements across the entire product assortment. I show mark-ups remained relatively stable between 2020 and 2023 across all major product categories. I document revenue-weighted mean mark-ups averaging $\sim 8.5\%$, with a peak in the third quarter of 2022 ($\sim 10.5\%$) - levels below those previously documented (see e.g. Dopper, MacKay, Miller, and Stiebale, 2024). Benchmarking against U.S. grocery survey margins on a four-quarter moving average basis reveals near-identical timing and volatility. Robust measures of mark-up distribution over the sample period show that changes are statistically insignificant except for a $\sim 7.4\%$ rise among top-percentile products, which seems to start to some extent two quarters prior, coincident with the CPI-price-index decoupling. Importantly, the 2022 increase is not driven by supermarket heterogeneity. To study the drivers of this late mark-up rise, I study transitions between mark-up percentiles, finding high persistence with mild mean reversion, but strongly weakened persistence in mid-to-late 2022. Analysis of fixed-basket mark-up indices further indicate reallocation of expenses by the third quarter of 2022. This is consistent with shifting tastes towards, and emergence of, new high-mark-up goods at about the same time that supply chain disruptions post-pandemic occurred. By documenting the evolution and distribution of mark-ups across the full product assortment, the paper provides evidence on whether and when recent price pressures and changing consumer preferences reflected broad-based margin expansion or reallocation toward higher-mark-up products.\\

Developing an approach to structural modelling and estimation of continuous demand systems opens the door to studying many other important questions which researchers have struggled to investigate with existing tools. How do discounts affect complementary goods? How do excise taxes impact staple good demand? How do private labels substitute higher-priced items cross-categories? These are questions of substantial policy relevance, the answers to which this paper may facilitate.\\

The rest of this paper is structured as follows. Section 2 reviews the relevant literature. Section 3 introduces the theoretical framework, the model setup, and examines aggregation through a representative agent. Section 4 introduces the data. Section 5 describes our application to study market power from intra-supermarket competition, and the empirical specification and identification strategy. Section 6 analyses the results. Section 7 concludes.\newpage

\section{Literature Review}

This paper's approach is in the tradition of continuous choice models such as the Rotterdam model (Theil, 1965), the translog model (Christensen, Jorgenson, and Lau, 1975), the Almost Ideal Demand System (Deaton and Muellbauer, 1980), and the multi-level demand system (Hausman, Leonard, and Zona, 1994). I adopt a tractable linear-demand framework, derived from a quasi-linear quadratic utility (see e.g. Amir, Erickson, and Jin, 2017; Choné and Linnemer, 2020). Linear demand remains a workhorse model in economic theory research due to straightforward intuition and application, closed-form solutions, and parsimony. However, it has not seen significant use in structural demand estimation. Pinkse, Slade, and Brett (2002), Pinkse and Slade (2004), and Lewbel and Nesheim (2019) are meaningful exceptions.\\


In industrial organisation, discrete-choice models such as the logit (McFadden, 1973) and the random-coefficients mixed logit (Berry, Levinsohn, and Pakes, 1995; Nevo, 2001) have been more popular. Previous related attempts at demand estimation across product categories in supermarkets in this tradition include Thomassen, Smith, Seiler, and Schiraldi (2017), and Dopper, MacKay, Miller, and Stiebale (2024). The present paper has been motivated by difficulties with adapting mixed logit to this setting, as part of a recent strain of research developed to address these: scaling to large choice sets (Armstrong, 2016; Lanier, Large, and Quah, 2023), incorporating flexible interaction effects and complementarities between goods (Berry and Haile, 2014; Fosgerau, Monardo, and de Palma, 2024; Monardo, 2025; Ershov, Laliberté, Marcoux, and Orr, 2025), and allowing choice sets inclusive of both individual and bundled alternatives (Gentzkow, 2007; Iaria and Wang, 2019; Sun, 2024). The present paper also speaks to quantitative marketing research on shopping basket models (see e.g. Manchanda, Ansari, and Gupta, 1999; Fox and Lazzati, 2017; Allen and Rehbeck, 2022).\\

My work also contributes to research on consideration sets, a relaxation of the assumption that individuals consider all feasible choices (Wright and Barbour, 1977). Many reasons have been given as to why the set of available choices may sometimes differ from a consumer's consideration set: behavioural heuristics (Hauser, 2014), cognitive limitations (Simon, 1957), bounded rationality and seller persuasion (Eliaz and Spiegler, 2011), imperfect information (Sovinsky, 2008), search (Caplin, Dean, and Martin, 2011), or evaluation costs broadly speaking (Hauser and Wernerfelt, 1990). Crawford, Griffith, and Iaria (2021) surveys how consideration set structure assumptions can impact demand estimation.\\

Lastly, the present paper contributes to a line of research on product differentiation as a function of distance between goods in a product variety space. This product variety space can be understood as a network, with goods as nodes and the distance between said nodes depending on cross-price effects (Ushchev and Zenou, 2018). Relevant papers include Hotelling (1929), Salop (1979), and Chen and Riordan (2007), which view product differentiation as a matter of location in an abstract space; Hoberg and Phillips (2016) and Pellegrino (2025), which measure distance in terms of cosine similarity between product characteristics; and Bajari et al. (2025) and Magnolfi, McClure, and Sorensen (2025), which use embeddings - low-dimensional representations of a latent product space - to define product's locations relative to every product pair.\newpage

\section{Theoretical Framework}
\subsection{Technical Setup}

Let there be $N$ consumers, indexed $i=1,\ldots,N$, and a finite number of non-numeraire goods $K$, indexed $k=1,\ldots,K$. A consumption bundle $\boldsymbol{q}_i\in \mathbb{R}^K$ denotes the quantities consumed by $i$ of each of the $K$ goods. Let $q_{i0}\in\mathbb R$ denote consumption of a numeraire good whose price is normalised to one. Assume that consumers hold continuous and monotonic rational preferences such that these admit a continuous utility function $U_i(\boldsymbol{q}_i, q_0),\ \forall i$, over alternative choices of $(\boldsymbol q_i,q_{i0})$. We assume this utility function to be twice-differentiable, and admit a
quasilinear representation of the form:

\begin{equation}
\begin{aligned}
    U_i(\boldsymbol{q}_i,q_{i0})&=U_i(\boldsymbol{q}_i)-\phi_iq_{i0}
\end{aligned}
\end{equation}

for $-\phi>0$ the marginal utility of income and $U_i(\boldsymbol q_i)$ the sub-utility obtained from the $K$ non-numeraire goods. Consumer $i$ has income/wealth $Y_i$. They spend part of it on the $K$ non-numeraire goods, and the remaining on the numeraire good. Since the numeraire price is normalised to one, the quantity of the numeraire equals the amount of money left over, i.e. the following budget constraint:

\begin{equation}
    q_{i0}=Y_i-\boldsymbol{q}_i'\boldsymbol{p}
\end{equation}

for $\boldsymbol{p}$ the price vector for the non-numeraire goods. The quasilinear specification implies that the numeraire enters utility linearly and separably.  Consequently, all curvature in preferences is captured by $U_i(\boldsymbol q_i)$, while the marginal utility of income $-\phi$ is constant.\\

For any such sub-utility function $U_i(\boldsymbol{q}_i)$, we can obtain a quadratic representation via second-order Taylor expansion around a reference bundle ($\overline{\boldsymbol{q}}$):

\begin{equation}\small
\begin{aligned}
    U_i(\boldsymbol{q}_i)\approx &\ U_i(\overline{\boldsymbol{q}})+(\boldsymbol{q}_i-\overline{\boldsymbol{q}})'\nabla U_i(\overline{\boldsymbol{q}})+\frac{1}{2}(\boldsymbol{q}_i-\overline{\boldsymbol{q}})'\nabla^2U_i(\overline{\boldsymbol{q}})(\boldsymbol{q}_i-\overline{\boldsymbol{q}})\\
    \approx &\ \boldsymbol{q}_i'\underbrace{(\nabla U_i(\overline{\boldsymbol{q}})-(\nabla^2U_i(\overline{\boldsymbol{q}}))\overline{\boldsymbol{q}})}_{\boldsymbol{\delta}_i,\ \text{marginal util. at reference bundle}}+\frac{1}{2}\boldsymbol{q}_i'\underbrace{(\nabla^2U_i(\overline{\boldsymbol{q}}))}_{-M_i,\ \text{Hessian}}\boldsymbol{q}_i+\underbrace{[U_i(\overline{\boldsymbol{q}})-\nabla U_i(\overline{\boldsymbol{q}})'\overline{\boldsymbol{q}}+\frac{1}{2}\overline{\boldsymbol{q}}'(\nabla^2U_i(\overline{\boldsymbol{q}}))\overline{\boldsymbol{q}}]}_{u_{i0},\ \text{utility level at reference bundle}}\\
    \approx &\ \boldsymbol{q}_i'\boldsymbol{\delta}_i-\frac{1}{2}\boldsymbol{q}_i'M_i\boldsymbol{q}_i+u_{i0}
\end{aligned}
\end{equation}

At $\overline{\boldsymbol{q}}=0$, $u_{i0}=U_i(0)$; we set this to zero across all $i$ without loss of generality (WLOG) for our utility maximisation exercise below. At $\overline{\boldsymbol{q}}=0$, we also have $\boldsymbol{\delta}_i=\nabla U_i(0)$, which can then be reinterpreted as the initial marginal utility per good for consumer $i$. The indirect utility function of said consumer is then:

\begin{equation}
\begin{aligned}
    v_i(\boldsymbol{p},Y_i)=&-\phi_i Y_i+\max_{\boldsymbol{q}_i}(U_i(\boldsymbol{q}_i)+\phi_i \boldsymbol{q}_i'\boldsymbol{p})\\
    =&-\phi_i Y_i +\frac{1}{2}(\boldsymbol{\delta}_i+\phi_i \boldsymbol{p})'M^{-1}_i(\boldsymbol{\delta}_i+\phi_i \boldsymbol{p})
\end{aligned}
\end{equation}

Applying Roy's identity:

\begin{equation}
\boldsymbol q_i(\boldsymbol p,Y_i)
=
-\frac{\nabla_{\boldsymbol p}v_i(\boldsymbol p,Y_i)}{\partial v_i(\boldsymbol p,Y_i)/\partial Y_i}= M^{-1}_i(\boldsymbol{\delta}_i+\phi_i \boldsymbol{p})
\end{equation}

In other words, quasilinear quadratic utility implies an exact linear demand, with zero income effects for the non-numeraire goods. Throughout this Chapter, I assume $M_i$ is symmetric and positive definite for all $i$. This guarantees the invertibility of $M_i$ and is necessary and sufficient for the well-behavedness of $U_i(\boldsymbol{q}_i),\ \forall i$, such that it admits a unique global maximum in $\boldsymbol{q}_i$ (Amir et al., 2017). Quasilinearity implies no income effects, a common simplification particularly in response to local changes in prices (Choné and Linnemer, 2020), and for goods individually representing a small portion of total income (Vives, 1987).  \\

A common assumption is to let, as we have so far, $\boldsymbol{q}_i\in\mathbb{R}^K$ - i.e. the consumption bundle can be any vector of $K$ non-negative real numbers. However, for $K$ goods, there are up to $2^K$ different feasible consumption bundles; for sufficiently large $K$, this assumption is therefore too strict. In this paper, I instead assume rational inattention; physical and technological conditions, mental capacity, and/or individual preferences are said to constrain the space of relevant bundles. In line with the literature, I call the subset of all feasible alternative consumption bundles that a consumer pays attention to and evaluates when making a decision the consumer's \textit{consideration set}.\\

Let $\mathcal{J}_i$ be the discrete set of these alternatives, or \textit{shopping baskets}, considered - individually or in linear combination - by consumer $i$ and indexed $j=1,\ldots,J_i$; and $A_{K\times J_i}$, the matrix representation of the consideration set: a sparse matrix of product presence per unique basket considered, each element $a_{kj}$ constituting the number of units of a given good $k$ in a given basket $j$ considered.\footnote{Each column can be thought of as the equivalent to one possible receipt during a purchase instance.} The columns of matrix $A_i$ are non-zero and unique, though not necessarily linearly independent.\\

Consumers decide which and how many alternatives to purchase. To account for the potential indivisibility of said alternatives, we can model the intensive and extensive margins of this choice via $\hat{\boldsymbol{z}}_{J_i\times 1}$, the unit-normalised choice bundle ($\boldsymbol{1}'\boldsymbol{\hat z}=1$); and $T_i$, the scalar number of transactions pursued. Both can be directly observed from transaction data.\\

\begin{figure}[H]
    \centering
    \caption{Example of shopping baskets in a consumer's consideration set}
    \label{fig:shopping-baskets-consideration-set}

    \vspace{0.25cm}

    \resizebox{0.75\textwidth}{!}{%
    $\displaystyle
    \renewcommand{\arraystretch}{1.15}
    \begin{array}{c@{\qquad}c@{\qquad\qquad}c}
    \boxed{
    \begin{array}{@{}l r@{}}
    \multicolumn{2}{c}{\texttt{ID: C236-491}}\\
    \multicolumn{2}{c}{\texttt{01/01/2026}}\\[-2pt]
    \multicolumn{2}{c}{\texttt{----------------}}\\
    \texttt{ITEM} & \texttt{QTY}\\
    \hline
    \texttt{Milk}   & 2\\
    \texttt{Bread}  & 1\\
    \texttt{Apples} & 6\\
    \texttt{Rice}   & 3\\
    \hline
    \multicolumn{2}{r}{\texttt{END RECEIPT}}
    \end{array}
    }
    &
    \boxed{
    \begin{array}{@{}l r@{}}
    \multicolumn{2}{c}{\texttt{ID: C431-667}}\\
    \multicolumn{2}{c}{\texttt{02/01/2026}}\\[-2pt]
    \multicolumn{2}{c}{\texttt{----------------}}\\
    \texttt{ITEM} & \texttt{QTY}\\
    \hline
    \texttt{Milk}   & 1\\
    \texttt{Bread}  & 2\\
    \texttt{Banana} & 4\\
    \texttt{Rice}   & 1\\
    \hline
    \multicolumn{2}{r}{\texttt{END RECEIPT}}
    \end{array}
    }
    &
    \Rightarrow \quad\qquad A_i =
    \left[
    \begin{array}{c@{\quad}c c@{\quad}c}
    \cdot & 2 & 1 & \cdot\\
    \cdot & 1 & 2 & \cdot\\
    \cdot & 6 & 0 & \cdot\\
    \cdot & 0 & 4 & \cdot\\
    \cdot & 3 & 1 & \cdot
    \end{array}
    \right]
    \end{array}
    $%
    }
\end{figure}\vspace{-1cm}
\bigskip\RaggedRight\singlespacing
\footnotesize \textbf{Notes}: Each consumer's consideration set incorporates all alternative shopping baskets considered by the consumer. Receipts of purchases made by a consumer are one way to identify the structure of said consumer's consideration set.
\normalsize
\justifying
\bigskip

We may thus express the goods consumption bundle $\boldsymbol{q}_i$ as a \textit{non-negative linear} function of a \textit{normalised} $\hat{\boldsymbol{z}}_i$:
\begin{equation}
    \boldsymbol{q}_i = T_iA_i\hat{\boldsymbol{z}}_i\qquad \& \qquad T_i,\hat{\boldsymbol{z}}_i\geq 0\qquad \& \qquad \boldsymbol{1}'\hat{\boldsymbol{z}}_i=1
\end{equation}

This condition summarises the constraints imposed by a consumer's consideration set on choice. When a consumer $i$ makes purchase decisions, the set of all feasible goods consumption bundles will span the \textit{conical hull} of matrix $A_i$:

\begin{equation}
\begin{aligned}
    &\boldsymbol{q}_i/T_i\in \mathrm{conv}(A_i)=\big\{\sum_{j=1}^J\boldsymbol{a}_{ij}x_j:\boldsymbol{a}_{ij}= A_i\boldsymbol{e}_j,x_j\in\mathbb{R}^K,\boldsymbol{1}'\boldsymbol{x}=1,j\in\mathcal{J}_i\big\}\\
    \Leftrightarrow
    &\ \boldsymbol{q}_i\in \bigcup_{T_i\geq0}T_i\cdot \mathrm{conv}(A_i)=\big\{\sum_{j=1}^J\boldsymbol{a}_{ij}x_j:\boldsymbol{a}_{ij}= A_i\boldsymbol{e}_j,x_j\in\mathbb{R}^K,j\in\mathcal{J}\big\}\\
    =&\ \mathrm{cone}(A_i)
\end{aligned}
\end{equation}

for $\mathrm{conv}(A_i)$ the convex hull implied by the $i$-th consideration set. Vector $\boldsymbol{e}_j$ is the $j$-th standard basis vector. Demand for goods per transaction depends on the proportion of choices that favour each shopping basket considered by the consumer. For $T_i\geq0$, one can scale those proportions up and down, as to get a cone. It is then trivial to see that whether separately optimising over the intensive and extensive margins or doing so jointly makes no difference for the choice of the optimal consumption bundle: no constraint is imposed.\\

Instead, consideration sets will play a role depending on the answer to the following question: when is $\mathrm{cone}(A_i)\subsetneq \mathbb{R}^K$?\\

\textit{\textbf{Proposition 1}: If $A_i$ is less than full row rank, $\boldsymbol{q}_i\in\mathrm{cone}(A_i)\subsetneq\mathbb{R}^K$. If and only if a standard basis vector $\boldsymbol{e}_i\notin\mathrm{cone}(A_i)$, then $\boldsymbol{q}_i\in\mathrm{cone}(A_i)\subsetneq\mathbb{R}^K$.}\hfill$\blacksquare$
\\

In other words, consumers' consideration sets bind if and only if not all goods are considered as standalone alternatives by a consumer. For the remainder of this Section, I study how this condition affects individual demand functions and verify if they are well-behaved.\\

We can write the consumer objective function's Lagrangean as:

\begin{equation}
\begin{aligned}
    \mathcal{L}(T_i,\hat{\boldsymbol{z}}_i)=& T_i\hat{\boldsymbol{z}}_i'A_i'\boldsymbol{\delta}_i  - \frac{1}{2} T_i^2\hat{\boldsymbol{z}}_i' (A_i' M_i A_i) \hat{\boldsymbol{z}}_i
- \phi_i \big(Y_i - T_i \boldsymbol{p}' A_i \hat{\boldsymbol{z}}_i\big)\\
&+ \underbrace{\boldsymbol{\hat\lambda}_i'}_{\text{non-negativity multiplier}} \hat{\boldsymbol{z}}_i +\underbrace{\xi_i}_{\text{unit-normalisation multiplier}}\boldsymbol{1}'\hat{\boldsymbol{z}}_i
\end{aligned}
\end{equation}

Maximising the Lagrangian of the consumer's basket choice problem in $\boldsymbol{\hat z}_i$, we obtain the following Karush-Kuhn-Tucker (KKT) conditions:

\begin{equation}\label{z_FOC}
    T_i(A_i'M_iA_i)\hat{\boldsymbol{z}}_i=A_i'(\boldsymbol{\delta}+\phi_i\boldsymbol{p})+\frac{1}{T_i}(\boldsymbol{\hat\lambda_i}+\xi_i\boldsymbol{1})\quad\boldsymbol{\hat\lambda}_i\geq0\quad\boldsymbol{\hat{z}}_i\geq0\qquad \boldsymbol{\hat\lambda}\circ\boldsymbol{\hat{z}}_i=\boldsymbol{0}
\end{equation}

Expressing the KKT stationarity condition in terms of $\boldsymbol{\hat z}_i$ introduces the first problem derived from our dimensionality constraint. \\

\textit{\textbf{Lemma 1}: $A_i'M_iA_i$ is positive semi-definite, and therefore not generally invertible.}\footnote{Proofs of all Lemmas may be found in the Appendix.}\\

To proceed, I multiply both sides of said condition by $(A_i'M_iA_i)^+$, for $(\cdot)^+$ the Moore-Penrose pseudoinverse (Moore, 1920; Penrose, 1955). However, on the LHS, $(A_i'M_iA_i)^+(A_i'M_iA_i)\neq I$ does not hold in general. A general demand function relative to $\boldsymbol{\hat z}_i$ must account for the so-called \textit{homogeneous part} of differential equations:

\begin{equation}
    T_i\boldsymbol{\hat{z}}_i=(A_i'M_iA_i)^+\big[A_i'(\boldsymbol{\delta_i}+\phi_i \boldsymbol{p}) +\frac{1}{T_i}(\boldsymbol{\hat\lambda_i}+\xi_i\boldsymbol{1})\big]+ (I-(A_i'M_iA_i)^+(A_i'M_iA_i))\boldsymbol{y}
\end{equation}

for any $\boldsymbol{y}\in \mathbb{R}^J$.\footnote{One may verify the propriety of this operation by noting that $(A_i'M_iA_i)(I-(A_i'M_iA_i)^+(A_i'M_iA_i))=0$, and $(A_i'M_iA_i)(A_i'M_iA_i)^+A_i'(\boldsymbol{\delta_i}+\phi_i \boldsymbol{p})=A_i'(\boldsymbol{\delta_i}+\phi_i \boldsymbol{p})$. The latter follows from the properties of the pseudo-inverse, as both $A_i'M_iA_i$ and $A_i'(\boldsymbol{\delta_i}+\phi_i \boldsymbol{p})$ are in the column space of $A_i'$.} There are thus a multiplicity of mappings of $\boldsymbol{p}$, as the strict concavity condition we had imposed on $M_i$ no longer holds for $A_i'M_iA_i$. In other words, the utility-maximising $\boldsymbol{\hat z}_i(\boldsymbol{p})$ is non-unique. Nonetheless,\\

\textit{\textbf{Proposition 2}: For the set of non-unique $\boldsymbol{\hat{z}}_i$, there is a unique $\boldsymbol{q}_i(\boldsymbol{p})=T_iA_i\boldsymbol{\hat{z}}_i$}:

\begin{equation} \label{demandsystem}
    \begin{aligned}
    \boldsymbol{q}_i(\boldsymbol{p})=A_i(A_i'M_iA_i)^+ \big[A_i'(\boldsymbol{\delta_i}+\phi_i \boldsymbol{p}) +\frac{1}{T_i}\boldsymbol{\hat\lambda_i}\big]
    \end{aligned}
\end{equation}\hfill$\blacksquare$

In short: despite the consideration set breaking the strict concavity assumption generally required for the well-behavedness of the utility function, its ensuing use (following the pseudoinverse and resulting non-unique $\boldsymbol{\hat{z}}_i$) ultimately isolates the single unique mapping from $\boldsymbol{p}$ to $\boldsymbol{q}_i$. The utility-maximising $\boldsymbol{q}_i(\boldsymbol{p})$ is unique, even if $\boldsymbol{\hat{z}}_i(\boldsymbol{p})$ is not. The above also shows that we need not worry about the indivisibility of $\boldsymbol{\hat z}_i$, as it does not constrain at all; we can assume divisibility in $\boldsymbol{\hat{z}}_i$ WLOG, as allowing for separate optimisation of the intensive and extensive margin of shopping basket choice does not change optimal $\boldsymbol{q}_i$. Separately considering the number of transactions pursued per consumer may nonetheless be relevant to demand aggregation, which we will get to later.\\

For $S_i$ the latent unconstrained Slutsky matrix of consumer $i$'s demand function - which is to say, that which we obtain where the Lagrangean constraints do not bind - we have that:

\begin{equation} \label{eq:nocompl}
    S_{i,ab}=\frac{\partial log(q_{ia})}{\partial log(p_b)}=\phi_i\frac{p_b}{q_{ia}}\boldsymbol{a_{ia}}'(A_i'M_iA_i)^+\boldsymbol{a_{ib}} \lesseqgtr 0 ,\ \forall a,b\in\mathcal{K}
\end{equation}

The sparsity of the $A_i$ and $M_i$ matrices minimises individual product demand's reliance on the price vector of the entire product assortment (and vice-versa). Despite losing positive definiteness (and nice properties such as diagonal dominance) under the dimensionality constraint, the demand system is still well-behaved. To see this, note:\\

\textit{\textbf{Lemma 2}: $A_i(A_i'M_iA_i)^+A_i'$ is symmetric and positive semi-definite.}\hfill$\blacksquare$\\

Symmetry and positive semi-definiteness in $S_i$ means the (unconstrained) individual demand function is aggregate-monotonic, and therefore satisfies the Law of Demand (Amir et al., 2017; see also Hildebrand, 1983).\footnote{It can also be shown that all own-price effects are strictly negative, i.e. $S_{i,aa}<0,\ \forall a\in\mathcal{K}_i$, as $M_i$ is positive definite and $A_i$ has no zero-rows by construction.}\\

\subsection{The Representative Consumer}

In some settings, only market-level data is available, so researchers are compelled to work with aggregate demand. This raises a standard problem: aggregation may fail to discipline structural parameters, and aggregate parameters may not directly map to individual preferences in the consumer population. To resolve this, it is commonly assumed that (i) there exists a representative consumer whose utility-maximising demand for goods reproduces the observed aggregate demand system, and (ii) that the representative preference parameters can be tied to underlying consumer population preferences and are recoverable via estimation. Gorman (1953) provides conditions under which a representative consumer rationalisation is admissible for some degree of consumer heterogeneity.\\

However, those conditions abstain from how heterogeneity in consumer preferences implies different consumers have different reservation prices. And it is when individual consumers' consumption bundle choices face different binding non-negativity constraints that the representative consumer interpretation becomes especially delicate, even if each consumer's demand is linear within a given interior regime. Aggregation then generally combines demands from different regimes, so the resulting market demand need not be globally representable as the solution of a single representative consumer problem.\\

As such, a \textit{representative consumer rationalisation} is generally not obtained in linear demand; no link can usually be established between a representative consumer's preferences and those of individual consumers.\footnote{Choné and Linnemer (2020) argues this point explicitly in its discussion of heterogeneity in the quasilinear quadratic utility model.} This raises concerns for demand estimation, which aims to obtain structural preference parameters to analyse counterfactual scenarios. Without a rationalisation, counterfactual analysis using linear demand is inappropriate.\\

In this section, I consider if and when the model set up above admits a representative consumer rationalisation.\\

First, we need a common language for consideration sets across all consumers. Let $A_{K\times J}$, for $J>K$ be the aggregate consideration set, made up of the $J$ unique columns across all $A_i$, such that, by the Minkowski sum, $\mathrm{cone}(A)=\sum_{i=1}^N\mathrm{cone}(A_i)$. Furthermore, let there exist a matrix $R_i\in\{0,1\}^{J\times J_i},\ \forall i$, such that $R_{i,ab}=1$ if the $a$-th column of $A$ matches the $b$-th column of $A_i$, mapping out which baskets appear in the latter. Note then that $A_i=AR_i,\ \forall i$. Additionally, define $\boldsymbol{z}_i=R_i\boldsymbol{\hat{z}}_i$, with $R_i$ playing the role of re-dimensioning $\boldsymbol{\hat{z}}_i$ from $J_i\times 1$ to $J\times 1$, with zeros in the new dimensions. The notation for the re-dimensioned choice bundle is $\boldsymbol{z}_i$. Then:

\begin{equation}
    \boldsymbol{q}_i=T_iA_i\boldsymbol{\hat{z}}_i=T_iAR_i\boldsymbol{\hat{z}}_i=T_iA\boldsymbol{z}_i
\end{equation}

This allows us to re-write the utility-maximisation problem of each consumer in terms of $\boldsymbol{z}_i$. The re-dimensioning cannot imply consumers make transactions they do not consider, meaning it effectively imposes a new zero-purchase constraint on the optimisation: $(I_J-R_iR_i')\boldsymbol{z}_i=0,\ \forall i$. Lastly, a dimensionality adjustment is applied to $\boldsymbol{\tilde\lambda}$ such that $\boldsymbol{\hat\lambda}_i'\boldsymbol{\hat{z}}_i=\boldsymbol{\lambda}_i'\boldsymbol{z}_i,\ \forall i$. Thus, in the same way we refer to $\boldsymbol{\lambda}_i$ as the non-negativity Lagrangean multiplier, we may refer to $\boldsymbol{\gamma}_i$ as the \textit{zero-purchase} Lagrangean multiplier of consumer $i$'s utility-maximisation problem. We will refer to the representative consumer's multipliers as $\boldsymbol{\Lambda}$ and $\boldsymbol{\Gamma}$ respectively.\\

We are now ready for the following proposition:\\

\textit{\textbf{Proposition 3}: Let the $\mathrm{cone}(A_i)\subsetneq \mathbb{R}^K$. Assume there exists a symmetric positive definite matrix $M$ such that each consumer’s latent unconstrained Slutsky matrix is the $M$-restricted response on the consumer’s feasible directions}:

\begin{equation}
    S_i\boldsymbol{y}= \phi_iM^{-1}\boldsymbol{y},\quad \forall\boldsymbol{y}\in \mathrm{span}(A_i),\  \forall i=1,\ldots,N
\end{equation}

\textit{Then, the aggregate demand:}

\begin{equation}
\begin{aligned}
    \boldsymbol{Q}(\boldsymbol{p})=\sum_{i=1}^N\boldsymbol{q}_i(\boldsymbol{p})=A(A'MA)^+[A'(\sum_i\boldsymbol{\delta}_i+\phi \boldsymbol{p})+\boldsymbol{\Lambda}+\boldsymbol{\Gamma}]
\end{aligned}
\end{equation}

\textit{can be represented as the solution of a representative consumer's constrained maximisation problem in the same local functional class.}\hfill$\blacksquare$\\

In other words, a binding consideration set and an agreed-upon shape for demand slopes across the individual consumers which consider them - i.e. what goods are substitutes/complements and to what extent - allow aggregation of preference heterogeneity in consideration sets, marginal utility of income, and initial marginal product utility as if reflecting the preferences of a single representative consumer.\\

The proposition assumes that consumers share a demand slope object $M$ in the $\mathrm{span}(A_i)$, across all $i$ in their latent \textit{unconstrained} Slutsky matrix. "Unconstrained" is in reference to a setting where we hold fixed the inequalities that create kinks and corners, and focus only on the response in the linear feasible directions. This notion is helpful to isolate local effects away from wider considerations on the active set of consumers, which we will get to later. Heterogeneity in the individual unconstrained Slutsky matrices can still arise: either outside the set of goods defined in a given $A_i$ (consumers may have different preferences over goods they do not buy); the considerations sets themselves; or the scalar factor $\phi_i$. All remaining heterogeneity is found in the constraints, and therefore only appears for the constrained Slutsky matrix of each consumer.\\

This is a far stronger claim than previously made in the literature, and relies directly on the existence of consideration sets. All consumers, despite holding consideration sets with different dimensions, can have their optimisation problem be written using the same basket space. After stacking their first-order conditions to obtain an aggregate condition, the summed Kuhn–Tucker multipliers are shown to not automatically satisfy complementary slackness for a representative consumer. My contribution is to show that, because basket representations are not unique, you can reassign those multiplier terms into new aggregate multipliers that generate the same aggregate demand while restoring the representative consumer’s KKT conditions. Thus, a combination of a basket choice bundle and a multiplier term can always be found that satisfies the KKT conditions for any given aggregate demand, recovering rationalisability.\\

Uniquely, this proposition directly addresses three of Eaton and Lipsey (1989)'s "awkward facts" about differentiated products, which were then argued to not be satisfied by representative consumer models: (i) that consumers purchase only a small subset of the products that are available in a given market; (ii) that consumers often have an approximate agreement as to product differentiation and what products are or not close substitutes/complements; and (iii) that consumers nonetheless decide upon different consumption bundles even where there are no relevant differences in income. \textbf{Proposition 3} claims a representative consumer rationalisation under precisely these conditions.\\

\subsection{Discussion}

While this is sufficient to confirm the existence of a representative consumer rationalisation, it is not enough to allow for separable identification of preference parameters - for a general $A$, $\boldsymbol{\delta}=\sum_{i=1}^N\boldsymbol{\delta}_i$ is not uniquely identified. However, note that:\\

\textit{\textbf{Lemma 3}: If $A$ is full row rank, then $A(A'MA)^+A'=M^{-1}$.}\hfill$\blacksquare$\\

Then:

\begin{equation}
\begin{aligned}
    \boldsymbol{Q}(\boldsymbol{p})=&A(A'MA)^+[A'(\sum_i\boldsymbol{\delta}_i+\phi \boldsymbol{p})+\boldsymbol{\Lambda}+\boldsymbol{\Gamma}]\\
    =&M^{-1}(\boldsymbol{\delta}+\phi \boldsymbol{p})+A(A'MA)^+[\boldsymbol{\Lambda}+\boldsymbol{\Gamma}]\\
\end{aligned}
\end{equation}

for $\boldsymbol{\delta}=\sum_i\boldsymbol{\delta}_i$ and $\phi=\sum_{i=1}^N \phi_i$.\\

In the Appendix, along with all proofs, I provide a full equilibrium analysis under Bertrand price competition, where $A$ is assumed less-than-full-row rank. I also run comparative statics under both full and less-than-full row rank.\\

However, outside of settings where consumers consider more products across shopping baskets than the total number of shopping basket alternatives considered, such a rank restriction is remarkably unlikely to hold. Even if linear dependence is the case for a single individual's consideration set - e.g. if individual purchase patterns are driven by a small number of latent factors - aggregation into a representative consumer will likely remove linear dependence in the aggregate consideration set. Each individual might even be influenced by a similar set of factors; but an aggregated consideration set with contributions from many individuals with differing factor loadings will have increasing effective diversity (or rank). In other words, while each individual's consideration set might be low rank, averaging over many individuals could add enough independent variation that the aggregate no longer possesses said dependencies. The answer to this question will be data-dependent, which we consider next.\newpage

\section{Data}

Point-of-sale data was obtained from \textit{SONAE MC}, a leading supermarket chain in the Portuguese food retail market. Information on transactions was collected through the stores' scanners in the moment of the sale. The dataset represents roughly 5\% of all transactions conducted across four supermarkets between February 2020 to February 2023. It covers date and location of each transaction; the set of products purchased per transaction both at the product level and across 4 different product category layers; and units purchased, gross sales value, and discounts issued per product per transaction. \\

The four supermarkets - Supermarket Maia, Hypermarket Mafra, Hypermarket Évora, and Hypermarket Portimão - were selected prior to the data analyses out of a list of 353 shops given their geographical distribution, distance from competitors, and size heterogeneity. The selection along these lines was made to match the model's assumptions - the primacy of intra-supermarket competition in particular - while including sufficient heterogeneity to consider the internal validity of the model's outcomes under different conditions. An evaluation of supermarket participation rates - i.e. the transaction volumes observed each period per supermarket - suggests that, for our sample of grocery stores, inter-supermarket competition is not a major concern (details are in the Appendix).\\ 

\begin{figure}
    \caption{Shop locations}
    \centering
    \includegraphics[scale=0.2]{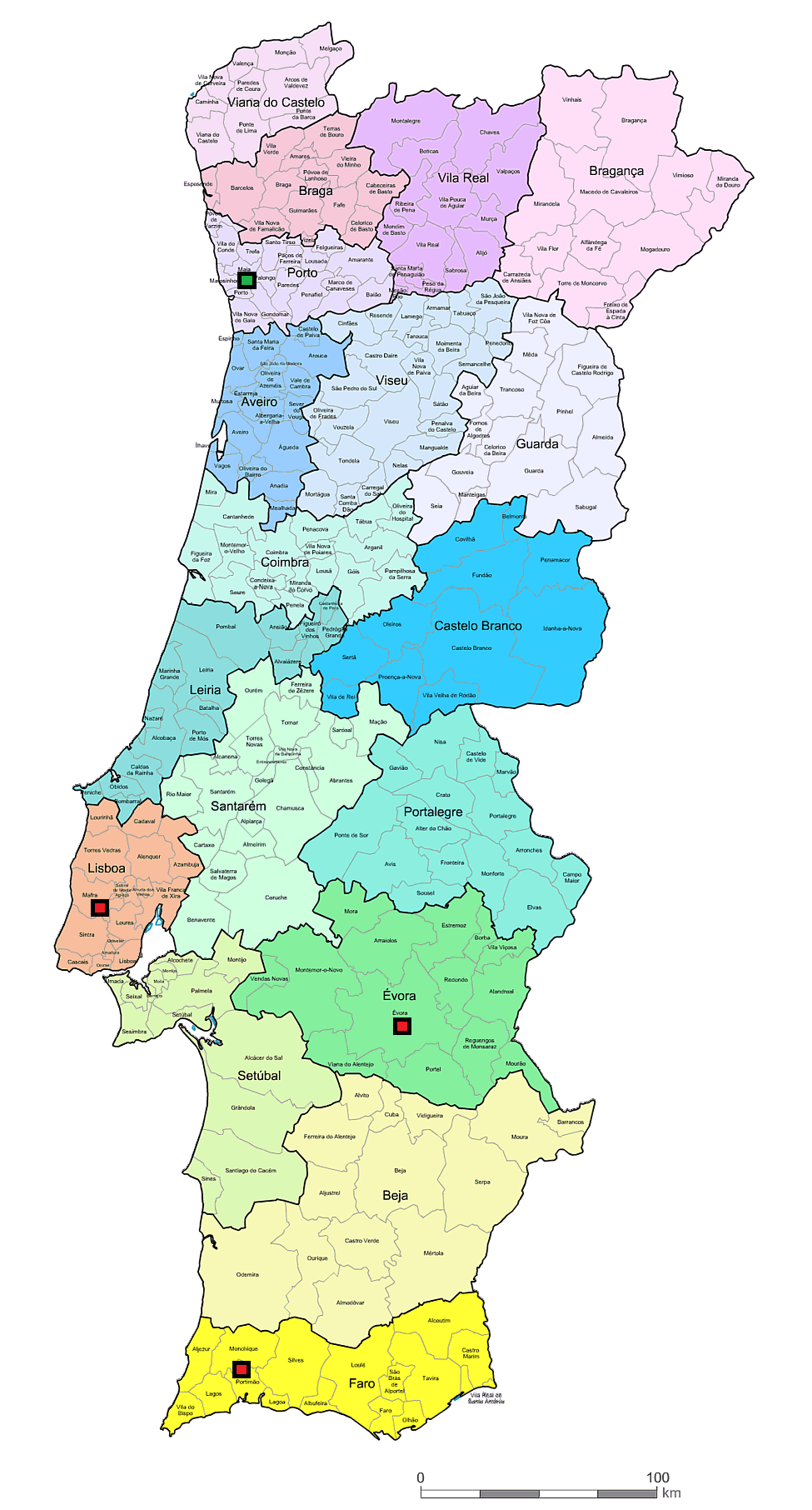}
    \\\RaggedRight\singlespacing
\footnotesize Notes: Hypermarket (red), supermarket (green). Hypermarkets are major retail hubs, meant to attract consumers to a distinct location within the urban fabric. Supermarkets are mid-sized stores targeting proximity to consumers.\\
\end{figure}

\normalsize\justifying

\renewcommand{\arraystretch}{0.7}
\begin{table}[H]
\caption{Transaction statistics across products and categories}
\label{basket1}
\centering
\begin{tabular}{lcc}
\toprule
& \textbf{Products} & \textbf{Categories} \\
\midrule 
A: Transaction size (number of items) & \\
Mean & 60.43 & 5.12 \\
Standard deviation & 67.99 & 3.69 \\
\midrule 
B: Percentile transaction size & \\
25th & 16 & 2 \\
Median & 36 & 4 \\
75th & 78 & 7 \\
\midrule 
C: Transaction share by size class & \\
$\leq$ 20 unique items / $\leq$ 2 groups & 32.97\% & 31.04\% \\
21-80 unique items / 3-8 groups & 42.80\% & 50.32\% \\
81-140 unique items / 9-14 groups & 13.74\% & 16.97\% \\
>140 unique items / >14 groups & 10.49\% & 1.67\% \\
\midrule
D: Revenue share by transaction size class & \\
$\leq$ 20 unique items / $\leq$ 2 groups & 8.04\% & 8.17\% \\
21-80 unique items / 3-8 groups & 30.52\% & 40.91\% \\
81-140 unique items / 9-14 groups & 23.21\% & 42.56\% \\
>140 unique items / >14 groups & 38.23\% & 8.36\% \\
\bottomrule 
\end{tabular}
\end{table}

\begin{table}[H]
\label{basket2}
\caption{Transaction statistics per store}
\centering
\begin{tabular}{lcc}
\toprule
\textbf{Shops by Type} & \textbf{Mean transaction size} & \textbf{SD transaction size} \\
\midrule
A: Hypermarket & & \\
Évora & 70.85 & 77.90 \\
Mafra & 52.68 & 58.12 \\
Portimão & 60.52 & 68.50 \\
\midrule
B: Supermarket & & \\
Maia & 49.00 & 51.35 \\
\bottomrule
\end{tabular}
\end{table}

\begin{table}[H] \label{Product1} 
\centering \caption{Product availability in stores and the average revenue brought by products at each level of availability} 
\begin{tabular}{lcccc}
\toprule
& \textbf{Products} & \textbf{Avg Revenue} & \textbf{Categories} & \textbf{Avg Revenue} \\ 
\midrule
At only 1 store & 2\% & \euro~ 2 206.4 & 5\% & \euro~ 1 334.0\\ At 2 stores & 3\% & \euro~ 693.5 & - & -\\ At 3 stores & 19\% & \euro~ 487.3 & 8\% & \euro~ 1 308.0\\ At all stores & 76\% & \euro~ 970.8 & 86\% & \euro~ 644 261.5\\ 
\bottomrule 
\end{tabular} 
\end{table}

\begin{table}[H]
\label{Product2}
\caption{Shop statistics}
\centering
\begin{tabular}{lcccc}
\toprule
 & \textbf{\#} & \textbf{\% Unique} & \textbf{\% Revenue} & \textbf{\% Unique Revenue} \\
\midrule
A: Hypermarket & & & & \\
Évora & 24 069 & 0.41\% & 31.91\% & 0.56\%\\
Mafra & 23 896 & 0.11\% & 13.73\% & 0.04\%\\
Portimão & 24 564 & 0.97\% & 44.00\% & 0.44\%\\
\midrule
B: Supermarket & & & & \\
Maia & 19 442 & 0.33\% & 10.36\% & 0.11\%\\
\bottomrule
\end{tabular}\\
\vspace{-0.5cm}\bigskip\RaggedRight\singlespacing
\footnotesize Notes: The stores share 18910 products.
\end{table}
\normalsize\justifying

The following procedures were applied to filter outliers and products for which insufficient data was made available. Any products with (i) fewer than 100 transactions, (ii) sold at zero price, and/or (iii) without any price variation were removed. Furthermore, product without close competitors in a given quarter across all supermarkets were also discarded. Lastly, any products which went unobserved in any quarter in any supermarket in the sample were assumed to not have been available for purchase in that quarter.\\

\begin{table}[H]
\label{Cat}
\caption{Product category statistics}
\centering
\begin{tabular}{lccc}
\toprule
\textbf{Categories} & \textbf{Products} & \textbf{Store presence} & \textbf{Sales share}\\
\midrule
Groceries (Salty) & 1 766 & 4 & 4.94\% \\
Groceries (Sweet) & 2 228 & 4 & 4.64\% \\
Soft Drinks & 1 267 & 4 & 7.09\% \\
Crates & 9 & 4 & 0.01\% \\
Hygiene & 1 597 & 4 & 4.21\% \\
Home Cleaning & 2 173 & 4 & 7.18\% \\
Frozen Goods & 758 & 4 & 3.27\% \\
Dairy and Derivates & 1 478 & 4 & 7.90\% \\
Beauty & 1 665 & 4 & 2.94\% \\
Essential Goods & 690 & 4 & 3.24\% \\
Butchery & 452 & 4 & 6.35\% \\
Fishmonger & 558 & 4 & 6.47\% \\
Charcuterie and Cheese & 1 182 & 4 & 6.85\% \\
Breakfast & 1 564 & 4 & 5.55\% \\
Fruits and Vegetables & 1 037 & 4 & 10.29\% \\
Bakery & 999 & 4 & 4.35\% \\
Wine and Spirits & 1 019 & 4 & 4.83\% \\
Take Away & 468 & 4 & 2.30\% \\
Luggage and Sport & 1 & 4 & 0.00\% \\
Bio and Healthy & 6 & 4 & 0.00\% \\
Newsagents & 884 & 4 & 0.96\% \\
DIY & 744 & 4 & 1.14\% \\
Petfood and Care & 798 & 4 & 2.15\% \\
Home Goods & 1 250 & 4 & 1.75\% \\
Stationery/Marketing & 80 & 4 & 0.18\% \\
Women Apparel & 1 & 3 & 0.00\% \\
Adult Non-Apparel & 2 & 1 & 0.00\% \\
Nursery  & 13 & 4 & 0.01\% \\
Large Domestic Goods & 1 & 3 & 0.01\% \\
Small Domestic Goods & 12 & 4 & 0.03\% \\
Others & 223 & 4 & 1.32\% \\
\midrule
TOTAL & 24 925 & & \\
\bottomrule
\end{tabular}
\end{table}

These corrections left us with 24 925 goods remaining, spanning 37 major categories, 118 sub-categories, and 571 sub-sub-categories. Each product is observed on average 1 210 times over the sample, with 175, 369, and 937 observations at the 25th, median, and 75th percentiles, respectively.\\

We observe 456 439 unique transactions - defined at the unit-per-product level - out of 499 047 baskets in total. The average number of observations per unique basket is 1.093; 2.16\% of unique baskets are observed more than once, accounting for 10.51\% of all baskets. Consistent with the shop-level results, hypermarkets exhibit larger and more dispersed basket sizes, stock more staple and unique items, and command a disproportionate revenue share. Aggregate gross sales equal \euro~82 636 508. Discounts total \euro~156 601, representing 0.19\% of gross sales.\\

\begin{figure}[H]
    \centering
    \caption{Revenue-weighted mean prices}
    \includegraphics[scale=0.7]{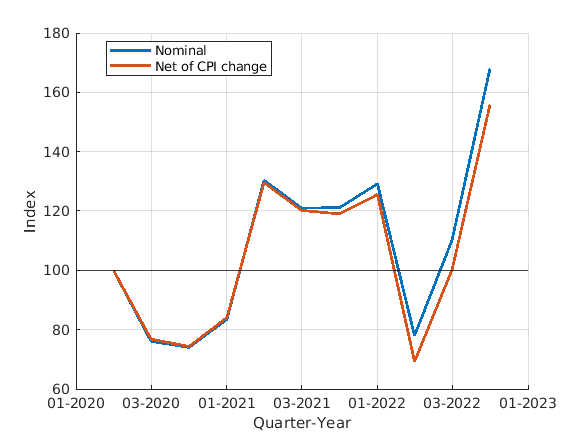}
    \label{pricesovertime}
\end{figure}

The revenue-weighted price index over the sample period is provided in Figure \ref{pricesovertime}, both in nominal and CPI-adjusted real terms.\\

The substantial observed volatility tracks the timings of the four COVID waves in Portugal, with the price peaks generally following a given wave: the first starting in Q2 2020; the second in Q4 2020; the third in Q4 2021, and the fourth in Q2 2022.\footnote{This information was obtained from the WHO COVID dashboard at data.who.int.} Sample volatility is significantly greater than that of the CPI, the latter with only a late increase driven by energy prices and supply chain disruption, with global container freight indices - a benchmark for the average cost of shipping containers across major world trade routes - peaking at the tail-end of 2021\footnote{See e.g. Monthly composite China Ningbo Container Freight Index at statista.com.} and crude oil prices peaking during 2022.\footnote{This information was obtained from statista.com.}\\

In this sample, 45.8\% of all goods are not purchased standalone. In what follows, I assume that this matches the structure of the true, latent, consideration set of the set of shoppers in the sample.\\

The aggregate consideration set is found to be full row rank. In other words, the application's setting reflects that where only non-negativity binds. As discussed in Chapter 1, this has two implications: all concerns raised by consideration sets and choice over combinations of goods will pertain primarily to the empirical implementation, rather than aforementioned theoretical implications; and aggregate preference parameters will be uniquely defined. Nonetheless, the Appendix includes a brief discussion on how to handle less-than-full-row-rank consideration sets in empirical specifications where it may be necessary.\newpage

\section{Empirical Application}
\subsection{Empirical Specification}

We now have relevant conditions under which we can learn of underlying population preference parameters using only market-level information.\\

In pursuing structural linear demand estimation, I address two concerns raised by Gandhi and Nevo (2021). Firstly is a curse of dimensionality; in linear demand models, as the number of alternatives grows larger, the number of own- and cross-price effect parameters grows (proportional to) quadratically in these. Secondly is price endogeneity. Linear demand incorporates many prices beyond a product's own, so overcoming price endogeneity concerns which arise from the simultaneity of demand and supply functions would usually require many exogenous and non-collinear instruments to separately identify price effects. A further, less discussed issue with linear aggregate demand estimation is the adjustment for individual constraints and corner solutions. In fact, representative consumer models tend to proceed without accounting for even that representative consumer's non-negativity constraint. As I show below, these introduce their own unique source of price endogeneity.\\

For simplicity, I will address these issues in turn, starting from the unconstrained consumer choice model. Demand is modelled at the average transaction by a representative consumer per supermarket: $\boldsymbol{Q}_{mt}/T_{mt}=\boldsymbol{q}_{mt}$. This is so that the estimated parameters do not partly pick up transaction-volume changes, and instead reflect per-transaction values: $\tilde{\boldsymbol \delta}=\frac{1}{T}\sum_{n=1}^N\boldsymbol{\delta}$ and $\tilde \phi=\frac{1}{T}\sum_{n=1}^N\phi$. The tilde is omitted for notation simplicity going forward. Let product utility have the following linear structure in product characteristics: $\boldsymbol{\delta}=X\boldsymbol{\beta}+\boldsymbol{v}$, for $X=[\boldsymbol{x}_1\ldots \boldsymbol{x}_K]$ a $N$ by $K$ matrix whose elements determine how much of a given observable characteristic $k$ good $n$ has; $\boldsymbol{\beta}$ a vector of the representative consumer's preference weights towards each of $K$ observable characteristics; and $\boldsymbol{v}$ an independent, identically distributed, mean zero, unit variance, unobserved error pertaining to demand shocks and latent product characteristics. For a given supermarket $m$ at a quarter-year pair $t$, unconstrained demand $\boldsymbol{q}^{\dagger}$ is therefore as follows:

\begin{equation}\label{unconstrained}
    \boldsymbol{q}_{mt}^{\dagger}=M^{-1}_m (X_{m}\boldsymbol{\beta}+\phi\boldsymbol{p}_{mt})+\boldsymbol{u}_{mt}
\end{equation}

for $\boldsymbol{u}=\boldsymbol{\varepsilon}+M^{-1}\boldsymbol{v}$, $\boldsymbol{\varepsilon}$ referring to the regression error.\\

Matrix $M$ is unknown. The first step is to therefore tackle the curse of dimensionality. To achieve this, I simplify this expression further by defining a Neumann series approximation of the inverse of the latent Hessian matrix on a proxy of observables $W$. The series approximation is helpful both for model fitting but also to minimise numerical instability and computation concerns stemming from inversion.

\begin{equation}\label{approxxx}
    M = I-\alpha W\quad \Rightarrow \quad M^{-1}\approx g(W) = \sum_{l=0}^L\alpha_lW^l
\end{equation}

for $L$ the number of series expansion terms. The scaling parameter $\alpha$ is made to allow differences between polynomials, i.e., $\alpha_l\neq\alpha^l,\ \forall l$. \\

To proxy for $M$, I take advantage of the frequency with which products are considered for joint purchase. In the economics literature, this is an approach closest to that in Atalay et al. (2023) and with roots in stochastic choice (e.g. Manzini, Mariotti, and Ülkü, 2019). The proxy itself is drawn from Tian et al. (2021), who approach the matter of cross-price effect estimation from a network science perspective. The authors use point-of-sale data and estimate complementarity and substitution relationships by the frequency with which goods are jointly purchased, relative to what would otherwise be expected.\\

A modified approach to that used in Tian et al. (2021) follows. Complements are defined as goods jointly purchased more often than expected, with the degree correlated to the extent of this. Substitutes are goods that both share the same complements and are jointly purchased less often than expected, with the degree measured similarly. The sales data is converted into a product-transactions matrix $B$ with every (row) good matched to every (column) transaction it appears in. Note that "transactions" by Tian et al. (2021) refers to non-unique "baskets". Earlier we have expressed the size of the set of all unique baskets as $J$; let us set $T$ as the total number of observed transactions. From the product-transactions matrix $B$, I compute two similarity measures.\\

The first similarity matrix - equivalent to the \textit{original measure} in Tian et al. (2021) - represents how similar the purchase patterns are between two products, yielding a measure of complementarity:

\begin{equation}
    W^{(c)}=A^{(c)}\circ \cos(\theta_c)
\end{equation}

The term $\cos(\theta_c)$ is a matrix whose elements are defined as, for a pair of goods $a,b\in \mathcal{K}$:

\begin{equation}
    \cos(\theta_c)_{ab} = \frac{\Xi_{ab}}{\sqrt{\Xi_{aa}}\sqrt{\Xi_{bb}}}
\end{equation}

for $\Xi=(D^\mathcal{P})^{-1}B(D^\mathcal{T})^{-1}B'(D^\mathcal{P})^{-1}$, $D^\mathcal{P}$ a diagonal matrix of the product degrees (how many transactions per good), and $D^\mathcal{T}$ a diagonal matrix of the transaction degrees (how many goods per transaction). $D^\mathcal{P}$ normalises the product vectors, whereas $D^\mathcal{T}$ weighs the relative importance of each transaction as the inverse of transaction size - smaller shopping baskets in the consideration set are assumed to pertain to stronger ties between the included goods. The matrix $\Xi$ is therefore a weighted version of the product-transactions matrix $B$.\\

The matrix $\cos(\theta_c)$ is defined under the assumption that all goods are complementary. It must therefore be supplemented by an analysis of joint purchases to determine the degree to which their frequency is \textit{significantly} different from what would be expected of independent goods. To achieve this, we produce $A^{(c)}$, the matrix of potential complements, with elements equal to one in case of significance.\\

Significance is assessed under a statistical null model - the Bipartite Configuration Model (BiCM) in this case - whose goal is to create a randomised product-transactions matrix that preserves the observed degree sequence (i.e. the number of connections) for both products and transactions, while randomising all other structural properties. The BiCM is favoured in our setting as it takes the bipartite feature of the product-transactions matrix into account.\\

In the literature dedicated to incorporating combinations of goods (not just individual goods) as options within a discrete-choice setting, the key concern lies in separating co-purchases which result from preferences vs taste correlation. Correlated purchases can be the outcome of things like minimising trip-related costs or time and stock considerations. Preferences toward co-purchases meanwhile may come from a love for variety or complementarities derived from various setting mentioned above. Both the equation for $\cos(\theta_c)$ and the BiCM are meant to adjust for product popularity and transaction size precisely to minimise the influence of correlations. \\

For each product pair $a$ and $b$ the expected number of common transactions $\mu_{ab}$ is computed as:

\begin{equation}
    \mu_{ab}=\frac{d^\mathcal{P}_{a}d^\mathcal{P}_{b}(\frac{1}{T}\sum_{t=1}^T (d^\mathcal{T}_{t})^2-\frac{1}{T}\sum_{t=1}^T d^\mathcal{T}_{t})}{\frac{1}{K}(\sum_{t=1}^Td^\mathcal{P}_{t})^2}
\end{equation}


for $T$ the number of transactions, and $d^\mathcal{P}_{i}$ the $i$-th diagonal element of $D^\mathcal{P}$. Once we have $\mu$ we can compute the observed number of joint purchases $C=BB'$. Lastly, we compare $c_{ab}$ with $\mu_{ab}$ using the Poisson distribution, on the assumption that in a sparse setting the probability of any two products co-occurring in a transaction is small:

\begin{equation}
    A^{(c)}_{ab}=\begin{cases}
        1\ \text{if}\ 1-F_{ab}(c_{ab}-1)<\alpha_c\\
        0\ \text{otherwise}
    \end{cases}
\end{equation}

where $F_{ab}(y)$ is the Poisson CDF with mean $\mu_{ab}$ and $\alpha_c$ is a chosen significance threshold.\\

The second similarity matrix - equivalent to the \textit{original substitutability measure} in Tian et al. (2021) - represents how similar pairs of products' complements are, yielding a measure of substitutability:

\begin{equation}
    W^{(s)}=I\{(A^{(c)})'A^{(c)}>0\}\circ A^{(l)}\circ\cos(\theta_s)
\end{equation}

The expression $\cos(\theta_s)$ can be straightforwardly derived from $\cos(\theta_c)$. For every given product pair:

\begin{equation}
    \cos(\theta_s)_{ab} = \frac{(\cos(\theta_c)\cos(\theta_c)')_{ab}}{\sqrt{\sum_{k=1}^K\cos(\theta_c)_{ak}}\sqrt{\sum_{k=1}^K\cos(\theta_c)_{bk}}}
\end{equation}

The matrix $A^{(l)}$ behaves similarly to $A^{(c)}$. A pair of products is flagged as having significantly less joint purchases than would be otherwise expected of independent goods if:

\begin{equation}
    F_{ab}(c_{ab})<\alpha_l
\end{equation}

where $F_{ab}(y)$ is once again the Poisson CDF with mean $\mu_{ab}$ and $\alpha_l$ is a chosen significance threshold. Should this be the case, the expression $I\{(A^{(c)})'A^{(c)}>0\}$ is meant to identify the presence of shared complementary goods, and $\cos(\theta_s)$ provides the weights.\\

The following graphs and tables provide details on what complementarity matrix $W^{(c)}$ and substitution matrix $W^{(s)}$ imply regarding within- and cross-category interaction effects:
\\

\begin{table}[H]
\label{crosscategoryexposure}
\caption{Proxy details - top categories by cross-category exposure}
\centering
\begin{tabular}{lccc}
\toprule
\textbf{Substitutes} & & \textbf{Complements} & \\
\midrule
1 & Dairy and Derivates & 1& Bio and Healthy\\
2 & Charcuterie and Cheese & 2& Nursery\\
3 & Essential Goods & 3& Home goods\\
\bottomrule
\end{tabular}
\end{table}

\begin{table}[H]
\label{crosscategoryexposure}
\caption{Proxy details - top categories by own-category exposure}
\centering
\begin{tabular}{lccc}
\toprule
\textbf{Substitutes} & & \textbf{Complements} & \\
\midrule
1 & Fruit and Vegetables & 1& Bio and Healthy\\
2 & Essential Goods & 2& Marketing\\
3 & Butchery & 3& Nursery\\
\bottomrule
\end{tabular}
\end{table}

\begin{figure}[H]
    \centering
    \caption{Proxy heatmap of own- and cross-category substitution - $W^{(s)}$}
    \includegraphics[scale=0.4]{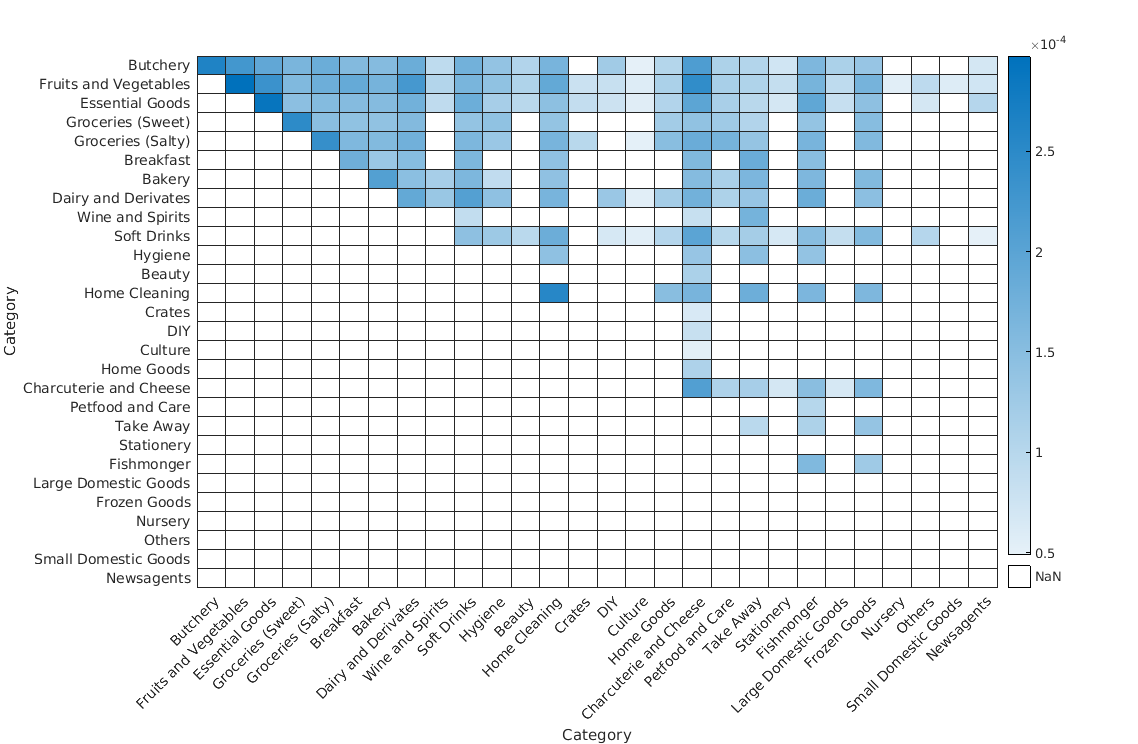}
    \label{heatmapsubs}
\end{figure}

\begin{figure}[H]
    \centering
    \caption{Proxy heatmap of own- and cross-category complementarity - $W^{(c)}$}
    \includegraphics[scale=0.4]{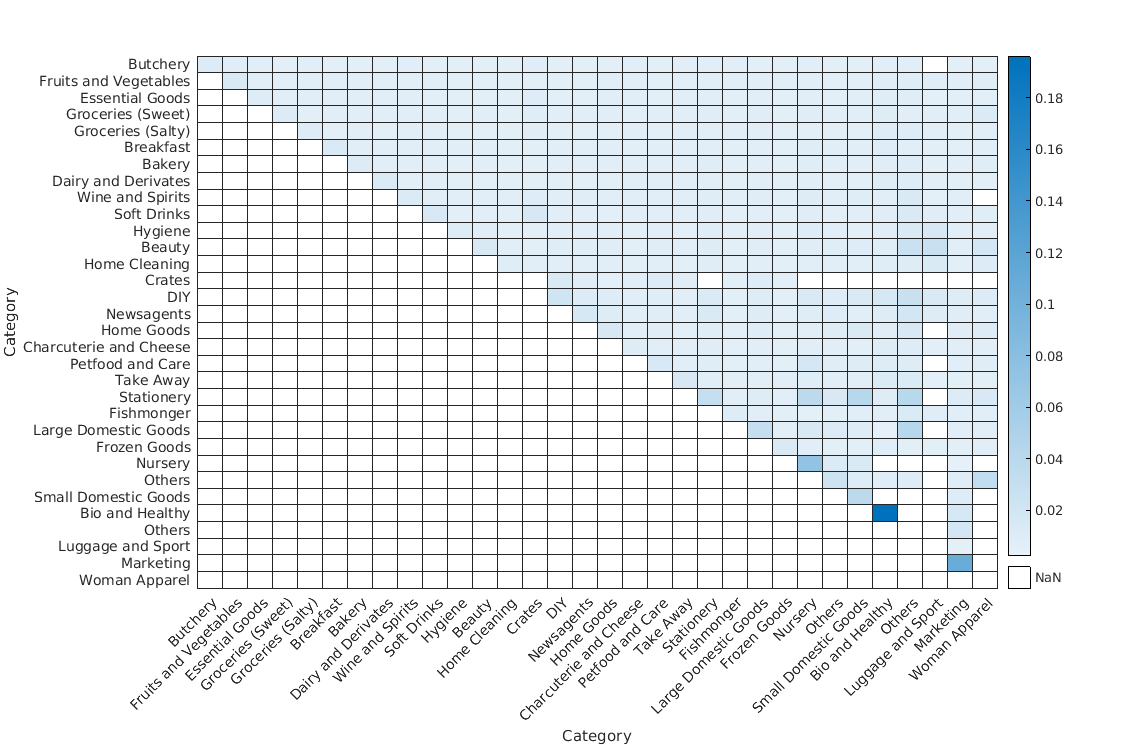}
    \label{heatmapcomps}
\end{figure}

The two similarity measures are tested both simultaneously and individually in the regression. Taking into account overfitting and multi-collinearity, the one providing the best fit is Tian et al. (2021)'s substitutability measure $W^{(s)}$ with two lags. All results below are provided relative to this specification.\\

Manzini, Mariotti, and Ülkü (2019) raise a pertinent criticism of attempts to define complementarity and substitution in terms of statistical regularities in co-purchase frequencies. For example, a pair of goods may appear as substitutes because they are rarely co-purchased, but the reason may be that consumers rarely consider them in the same shopping trip. If a product pair becomes more frequently co-purchased, an association-based criterion can still change depending on how marginal frequencies shift. Manzini, Mariotti, and Ülkü (2019) provide an example that reveals how higher joint consumption can reduce measured association under some normalisations.\\

These concerns are addressed to some degree by the adjustments made over product popularity and transaction size. The statistical nature of the approach itself requires a high degree of confidence in the strength of the implied linkage. Nonetheless, it is important to note that, while our proxy from co-purchase frequencies is used as a basis for likely demand interactions, it is not by itself assumed to identify model-free complementarity or substitution. Structural interpretation comes only after integration of the proxy onto the demand model and said model's parametric estimation.\\

The unconstrained linear demand model with the Neumann series approximation to the Hessian matrix inverse can be written as follows:

\begin{equation}
\begin{aligned}\label{spatecon}
    \boldsymbol{q}_{mt}^{\dagger} & \approx \sum_{l=0}^L\alpha_{l}W^l_m(X_m\boldsymbol{\beta}+\phi\boldsymbol{p}_{mt}) + \boldsymbol{u}_{mt}\\
    &=X_m\boldsymbol{\beta}+\phi\boldsymbol{p}_{mt}+\sum_{l=1}^L\alpha_{l}W^l_m(X_m\boldsymbol{\beta}+\phi\boldsymbol{p}_{mt}) + \boldsymbol{u}_{mt}
\end{aligned}
\end{equation}

The formulation in (\ref{spatecon}) shows how the original regressors $X\boldsymbol{\beta}$ and $\phi\boldsymbol{p}$ are augmented by the proxy lags $W^l(X_m\boldsymbol{\beta}+\phi\boldsymbol{p}_m)$ weighted by the coefficients $\alpha_l$. In practice, this means that our estimation model can be set up as a standard linear regression that includes both the original variables and their corresponding lagged versions. This is computationally favourable as it allows lags to be precomputed, and is effectively a series of linear combinations, which we may estimate via standard techniques. It also enables the separate identification of $\alpha_{l}$ from $\phi$ if so desired, e.g. via the delta method.\\


Stacking across supermarkets, and including covariate interactions with prices to account for coefficient heterogeneity, the final unconstrained model specification is as follows:

\begin{equation}\label{yeehawww}
        \boldsymbol{q}_{mt}^{\dagger} \approx X_m\boldsymbol{\beta}+(\overline{\phi}+ X_m\boldsymbol{\psi} )\circ\boldsymbol{p}_{mt}+\sum_{i=1}^M\boldsymbol{1}_{\{i=m\}}\bigg[\sum_{l=1}^L\alpha_{j}W^l_m(X_m\boldsymbol{\beta}+(\overline{\phi}+ X_m\boldsymbol{\psi})\circ\boldsymbol{p}_{mt})\bigg]+\boldsymbol{u}_{mt}
\end{equation}

The second step is to address price endogeneity and instrument collinearity. The proxy lags relieve much of the latter concern, as they allow just a single instrumental variable sufficient variation to cover all prices: $W_m^l(\overline\phi+X_m\psi)\circ\boldsymbol{p}_{mt}^{IV}$. However, a separate concern results: subsequent proxy lag interactions with covariates ($W_m^lX_m$) contain less new information, re-introducing multi-collinearity. To address this, a Principal Component Analysis (PCA) is applied to $X$ to orthogonalise covariates and extract the key factors driving variation in the data. Covariates are restricted to the smallest combination of components which contains 90\% of $X$ variation or higher. The top 10 principal components in variation are retained to form heterogeneous price coefficients. The use of PCA to address multi-collinearity is commonplace in analyses of multivariate and high-dimensional data (Koch, 2013). A slew of recent working papers in industrial organisation have attempted to use PCA to reduce the full set of characteristics (see e.g. Conlon, 2013; Backus, Conlon, and Sinkinson, 2021).\\

The last step is the handling of the Lagrangean constraints on $\boldsymbol{z}$. Consider the general, constrained form of (\ref{unconstrained}):

\begin{equation}
    \boldsymbol{q} = M^{-1}(\boldsymbol{\delta}+\phi\cdot\boldsymbol{p})+A(A'MA)^+(\boldsymbol{\Lambda}(\boldsymbol{\delta}, \boldsymbol{p})+\boldsymbol{\Gamma}(\boldsymbol{\delta}, \boldsymbol{p}))
\end{equation}

The term $A(A'MA)^+(\boldsymbol{\Lambda}+\boldsymbol{\Gamma})\neq0$ if, for any individual within the consumer population $\exists z_{ij}:\ z_{ij}=0$, i.e. if there are corner solutions either due to prices ($\boldsymbol{\Lambda}$) or due to consideration set heterogeneity ($\boldsymbol{\Gamma}$). Together, they create a wedge between $\boldsymbol{q}$ and the unconstrained model $\boldsymbol{q}^{\dagger}$: $\Delta\boldsymbol{q}=\boldsymbol{q}-\boldsymbol{q}^{\dagger}=A(A'MA)^+(\boldsymbol{\Lambda}+\boldsymbol{\Gamma})$. For demand estimation, this is problematic. Because the wedge is price-dependent, it introduces an additional form of endogeneity. Not only are $\boldsymbol{\Lambda}$ and $\boldsymbol{\Gamma}$ unobserved, but we also cannot instrument said endogeneity away - the wedge is a function of price, rather than being correlated with it.\\

A few approaches to the non-negativity problem and $\boldsymbol{\Lambda}$ have been suggested in the literature. The most common are censored regressions, such as Tobit models (e.g. Wales and Woodland, 1983; Lee and Pitt, 1986). Such an approach is inconsistent with utility maximisation, since non-negativity is introduced ex-post rather than assumed to be considered by consumers during utility maximisation. As far as I am aware, the economics literature does not discuss the matter of $\boldsymbol{\Gamma}$ and the need to account for corner solutions driven by consideration set heterogeneity at all. In our favour, however, is the representative consumer rationalisation argument made earlier. There, we proved that there are an infinite number of alternative $\boldsymbol{\Lambda}$ and $\boldsymbol{\Gamma}$ that can satisfy the KKT conditions on the aggregated FOCs of the consumer population as if it were that of a representative consumer. This means we need only find the one $\boldsymbol{z}$ that delivers the best $A(A'MA)^+(\boldsymbol{\Lambda}+\boldsymbol{\Gamma})$ fit to the model. \\

To do so, note the following:\\
\vspace{-1cm}

\begin{equation}\small
    \boldsymbol{Q} = M^{-1}(\boldsymbol{\delta}+\phi\boldsymbol{p})+A(A'MA)^{+}\lambda(\boldsymbol{p}) \quad \Leftrightarrow\quad \boldsymbol{Q} = A \cdot \arg\min_{\boldsymbol{z} \ge 0}||A\boldsymbol{z} - M^{-1}(\boldsymbol{\delta}+\phi \boldsymbol{p})||^2_M
\end{equation}

This expression is stated in terms of total demand. To move to $\boldsymbol{q}$, note:

\begin{equation}
    \boldsymbol q^\dagger=
    \frac{1}{T}M^{-1}(\boldsymbol\delta+\phi\boldsymbol p)=M^{-1}(\tilde{\boldsymbol\delta}+\tilde\phi\boldsymbol p)
\end{equation}

where $\tilde{\boldsymbol\delta}=\frac{1}{T}\sum_{i=1}^N\boldsymbol{\delta}_i$ and
$\tilde\phi=\frac{1}{T}\sum_{i=1}^N\phi_i$ are the structural parameters as mentioned before. For any $T>0$ and $\boldsymbol{\tilde z}=T\boldsymbol{z}$:

\begin{equation}
\frac{1}{T}
A\cdot
\arg\min_{\boldsymbol{\tilde z}\ge0}
\left\{
    \left\|A\boldsymbol{\tilde z}-
    M^{-1}(\boldsymbol\delta+\phi\boldsymbol p)
    \right\|_M^2
\right\}
=
A\cdot
\arg\min_{\boldsymbol z\ge0}
\left\{
    \left\|A\boldsymbol z-
    \boldsymbol q^\dagger
    \right\|_M^2
\right\}
\end{equation}

Since $\boldsymbol z\ge0$ if and only if $\tilde{\boldsymbol z}\ge0$, and since $A\boldsymbol z=(1/T)A\tilde{\boldsymbol z}$, scaling the target from $\boldsymbol Q^\dagger$ to $\boldsymbol Q^\dagger/T$ simply scales the non-unique NNLS-optimal coefficient vector $\boldsymbol{z}$. Thus the lower-case empirical object satisfies the analogous representation:

\begin{equation}
    \boldsymbol q = A\cdot \arg\min_{\boldsymbol z\ge0}
    \{||A\boldsymbol z- \boldsymbol q^\dagger||_M^2\}
\end{equation}

From here, I propose the following approach. First, compute $\boldsymbol{\hat q}^\dagger$, the fitted values of an initial, "naive", as-if-unconstrained regression and estimate $\hat{M^{-1}}$. Second, find $\boldsymbol{r}=\boldsymbol{\hat q}^\dagger-A\cdot\arg\min_{\boldsymbol{z}\geq0}\{||A\boldsymbol{z}-\boldsymbol{\hat q}^\dagger||^2_{\hat{M}}\}$. Third, after getting $\boldsymbol{r}$, plug the result back onto the LHS of the regression so the dependent variable becomes $\boldsymbol{q}^{observed}+\boldsymbol r$ (or $\boldsymbol q+\boldsymbol r/\mathrm{sd}_Q$ in standardised units), i.e., a control-function step. Lastly, re-run the regression, obtaining a new estimate for $\hat M$ and repeating the process again until convergence.\\

Notice how $\boldsymbol{z}$ is the choice variable, yet it plays no direct role unless multiplied by $A$. This means we can take advantage of its non-uniqueness to facilitate the solving of the weighted non-negative least squares problem without that affecting the overall estimation. Furthermore, the NNLS-optimal $\boldsymbol{z}$ term need not be constrained to $\boldsymbol{1}'\boldsymbol{z}=1$. Although $\boldsymbol{Q}/T$ admits a normalised transaction-share representation, which would need to be bounded in such a way, the $\boldsymbol{z}$ in the correction step is not necessarily that representation. It is meant to show that there exists some representative-consumer basket vector and multiplier configuration that rationalises aggregate demand. It is a non-unique object only constrained to satisfying the KKT conditions, of which non-negativity is one. Total mass $\boldsymbol{1}'\boldsymbol{z}$ is therefore not a transaction count. It is just the coefficient mass of the chosen representation. That mass has no direct structural interpretation unless additional restrictions are imposed.\\

I also considered two alternative specifications: $\boldsymbol{r}=\boldsymbol{\hat q}^\dagger-A\cdot\arg\min_{\boldsymbol{z}\geq0}\{||A\boldsymbol{z}-\boldsymbol{\hat q}^\dagger||^2_{\hat{M}}+\tau\boldsymbol{1}'\boldsymbol{z}\}$ and $\boldsymbol{r}=\boldsymbol{\hat q}^\dagger-A\cdot\arg\min_{\boldsymbol{z}\geq0}\{||A\boldsymbol{z}-\boldsymbol{\hat q}^\dagger||^2_{\hat{M}}+\tau||\boldsymbol{z}||\}$. Both follow the spirit of the $\boldsymbol{\Gamma}$, either by centralising demand on a few popular shopping baskets or by decentralising it to reflect consideration set heterogeneity. Likely due to the dimension of $A$, our results are robust to either specification. In other words, the primary consideration is the adjustment of the empirical model to the aggregated individual corner solutions which result from prices.

\subsection{Identification Strategy}

The estimation procedure itself is carried out via Two-Stage Least Squares estimation. I take advantage of the stream of observations on product purchases across five geographically distinct supermarkets each quarter of each year in our database, as well as the repeated cross-sectional nature of our data and our proxy lags, to define matrix $Z$ of price instruments. I compute the average price of a given product's competitors across other supermarkets as our primary price instrument (Hausman, 1996). As exogenous variables, also in $X$, I include intercepts, supermarket and quarter FEs, private-label brand FEs, product category FEs at three levels of detail, and the number of supermarkets in our sample that each good is present in every time period. Lastly, following Pinkse and Slade (2004), for every additional proxy lag included in the model, additional instruments $W^jZ$ are computed.\\

Our estimation procedure will be successful provided $\mathbb{E}(Z'(\boldsymbol{\varepsilon}_{mt}+M^{-1}\boldsymbol{v}))=0$. Insofar as $\boldsymbol{v}$ is mean zero, independent across agents, and is composed of short-term product utility shocks, it is unlikely to play a significant role - i.e. the assumption is then equivalent to the usual exogeneity condition on IVs. This would be less so the case if it were instead composed of latent product characteristics relevant to product competition. For the latter case, the rest of the paper assume exogenous characteristics over our period of analysis (see, e.g. Lee, 2024, for early positive indicators on this matter).\\

However, note that if $\boldsymbol{u}=M^{-1}\boldsymbol{v}+\boldsymbol{\varepsilon}$, then $Var(\boldsymbol{u})=\sigma^2M^{-2}$. Even if $\mathbb{E}(Z'(M^{-1}\boldsymbol{v}))=0$ and our estimates are unbiased, this variance has significant effects on the error terms which need to be accounted for separately. The problem is that we do not observe $M^{-1}$, but rather an estimated $\hat M^{-1}$. Therefore, while correction may be desirable, the use of $\hat M^{-1}$ may compound sampling error without much benefit. There is no guarantee the same proxy that is good for predicting cross-price effects in the conditional mean is also good for capturing the covariance of latent shocks. Instead, I cluster standard errors at the store-quarter level, allowing arbitrary covariance across all product observations in that store-quarter. That is exactly what we want if we suspect the residual has a complicated cross-product structure but also want to avoid over-committing to a particular parametric structure.\newpage

\section{Results}

The results of our empirical strategy are presented below. The first-stage results pertaining to the excluded instrumental variables are as follows:\\

\begin{table}[H]\centering
\caption{Standardised 2SLS first stage}
\label{tab:first_stage_excluded_instr}
\begin{threeparttable}
\footnotesize
\setlength{\tabcolsep}{10pt}
\begin{tabular}{@{} l
S[table-format=-1.2, table-space-text-post={***}]
S[table-format=1.2]
@{}}
\toprule
& \multicolumn{1}{c}{Coefficient}
& \multicolumn{1}{c}{SE} \\
\midrule
$PC\{IV,1\}$  & 4.94* & 2.73 \\
$PC\{IV,2\}$  & -0.15** & 0.07 \\
$PC\{IV,3\}$  & 0.62 & 2.76 \\
$PC\{X,1\}\circ PC\{IV,1\}$  & -0.01** & 0.00 \\
$PC\{X,2\}\circ PC\{IV,1\}$  & -0.03*** & 0.01 \\
$PC\{X,3\}\circ PC\{IV,1\}$  & 0.02** & 0.01 \\
$PC\{X,4\}\circ PC\{IV,1\}$  & -0.04*** & 0.01 \\
$PC\{X,5\}\circ PC\{IV,1\}$  & -0.04*** & 0.01 \\
$PC\{X,6\}\circ PC\{IV,1\}$  & -0.02* & 0.01 \\
$PC\{X,7\}\circ PC\{IV,1\}$ & 0.31*** & 0.10 \\
$PC\{X,8\}\circ PC\{IV,1\}$ & 0.05*** & 0.01 \\
$PC\{X,9\}\circ PC\{IV,1\}$ & 0.05*** & 0.01 \\
$PC\{X,10\}\circ PC\{IV,1\}$ & 0.00 & 0.01 \\
$PC\{X,1\}\circ PC\{IV,2\}$ & 0.01 & 0.01 \\
$PC\{X,2\}\circ PC\{IV,2\}$ & 0.01 & 0.05 \\
$PC\{X,3\}\circ PC\{IV,2\}$ & 0.05 & 0.05 \\
$PC\{X,4\}\circ PC\{IV,2\}$ & -0.01 & 0.01 \\
$PC\{X,5\}\circ PC\{IV,2\}$ & -0.03 & 0.03 \\
$PC\{X,6\}\circ PC\{IV,2\}$ & -0.04 & 0.03 \\
$PC\{X,7\}\circ PC\{IV,2\}$ & -0.04** & 0.02 \\
$PC\{X,8\}\circ PC\{IV,2\}$ & 0.00 & 0.02 \\
$PC\{X,9\}\circ PC\{IV,2\}$ & 0.06*** & 0.02 \\
$PC\{X,10\}\circ PC\{IV,2\}$ & 0.05*** & 0.02 \\
$PC\{X,1\}\circ PC\{IV,3\}$ & -0.05** & 0.02 \\
$PC\{X,2\}\circ PC\{IV,3\}$ & -0.01 & 0.01 \\
$PC\{X,3\}\circ PC\{IV,3\}$ & 0.03 & 0.02 \\
$PC\{X,4\}\circ PC\{IV,3\}$ & 0.05*** & 0.01 \\
$PC\{X,5\}\circ PC\{IV,3\}$ & 0.01 & 0.02 \\
$PC\{X,6\}\circ PC\{IV,3\}$ & 0.02 & 0.02 \\
$PC\{X,7\}\circ PC\{IV,3\}$ & -0.60** & 0.25 \\
$PC\{X,8\}\circ PC\{IV,3\}$ & -0.12*** & 0.04 \\
$PC\{X,9\}\circ PC\{IV,3\}$ & -0.02& 0.02 \\
$PC\{X,10\}\circ PC\{IV,3\}$ & -0.00 & 0.02 \\
\vspace{-0.25cm}\\
\bottomrule
\end{tabular}
\end{threeparttable}\\
\vspace{-0.5cm}\bigskip\RaggedRight\singlespacing
\footnotesize Notes: Coefficients and standard errors are multiplied by $10^2$. $^{***}p<0.01$, $^{**}p<0.05$, $^{*}p<0.1$.
\end{table}

On the one hand, the majority of significant effects are primarily observed amongst the interaction effects between the product characteristics and the Hausman instrument without proxy lags; on the other hand, two of the effects greatest in magnitude pertain to an interaction effect including the second proxy lag. To assess instrument informativeness and relevance across both the base and consideration-set-adjusted model, I obtained the following diagnostics:\\

\begin{table}[H]\centering
\caption{Model diagnostics and PCA summary}
\label{tab:model_diagnostics}
\begin{threeparttable}
\footnotesize
\setlength{\tabcolsep}{8pt}
\begin{tabular}{lcc}
\toprule
& \multicolumn{1}{c}{Base model}
& \multicolumn{1}{c}{Consideration-set-adjusted model} \\
\midrule
\textbf{PCA} \\
\quad \# PCs for 90\% variance & 76 & 76 \\
\quad Variance explained by top 10 PCs (\%) & 76.6 & 76.6 \\
\addlinespace
\textbf{Fit \& diagnostics} \\
\quad Adjusted $R^{2}$ & 0.630 & 0.564 \\
\quad F-test (122, 557514) & 42799*** & 862650*** \\
\quad Durbin-Wu-Hausman $\chi^{2}$ (33) & 142.74*** & 5565.4***\\
\bottomrule
\end{tabular}
\end{threeparttable}\\
\vspace{-0.5cm}\bigskip\RaggedRight\singlespacing
\footnotesize Notes: Robust standard errors in parentheses. $^{***}:p<0.01$, $^{**}:p<0.05$, $^{*}:p<0.1$. PCA is computed on the initial covariate set $X$. Degrees of freedom for the Wald test are shown as $(\text{df1}, \text{df2})$; the Durbin-Wu-Hausman statistic displays the $\chi^{2}$ degrees of freedom in parentheses.
\end{table}

The PCA applied to $X$ and its proxy lags to orthogonalise covariates and extract the key factors driving variation in the data reduces their number to $76$. The top 10 principal components are used to form heterogeneous price coefficients in both specifications, and are shown to explain up to 76.6\% of the variation across all covariates. Combined with quarter-year and supermarket FEs as well as the proxy lags, the total number of regressors is 122, over 557514 observations. The consideration-set-adjusted model converges after three iterations.\footnote{Convergence is assessed on whether the largest before-and-after entrywise change between either the vector of coefficients $\boldsymbol{\theta}$ or the wedge $\boldsymbol{r}$ relative to their previous maximimum element respectively is less than a pre-specified tolerance, namely 0.1. They are 0.0681 and 0.4766 respectively after the third iteration.}\\

The model retains substantial explanatory power even in its more restrictive version. The difference in fit is suggestive how much of the unadjusted results may be spurious. In each case, the included regressors are jointly informative, and the Hausman instrument - supported by its proxy lags - successfully minimises existing price endogeneity, which as expected further decreases once adjusting for corner solutions.\\

The coefficients pre- and post-NN adjustment are as follows, focusing on the parameters of greatest interest:

\begin{table}[H]\centering
\caption{Standardised 2SLS parameters of interest (Pre-NNLS vs Post-NNLS)}
\label{tab:2sls_params_pre_post_econ}
\begin{threeparttable}
\footnotesize
\setlength{\tabcolsep}{10pt}
\begin{tabular}{@{} l
S[table-format=2.4] S[table-format=2.4]
@{}}
\toprule
& \multicolumn{1}{c}{Base model} & \multicolumn{1}{c}{Consideration-set-adjusted model} \\
\midrule
$\bar{\phi}$   & 8.1426**    & 2.0504 \\
               & {\scriptsize(3.3137)} & {\scriptsize(3.2655)} \\
$\eta_{1}$     & -0.0223     & -0.1227*** \\
               & {\scriptsize(0.0180)} & {\scriptsize(0.0186)} \\
$\eta_{2}$     & -0.1305     & -0.6104*** \\
               & {\scriptsize(0.0859)} & {\scriptsize(0.0898)} \\
$\eta_{3}$     & -0.0856     & -0.5029*** \\
               & {\scriptsize(0.0766)} & {\scriptsize(0.0788)} \\
$\eta_{4}$     & -0.0046     & 0.0029 \\
               & {\scriptsize(0.0085)} & {\scriptsize(0.0085)} \\
$\eta_{5}$     & 0.0577      & 0.2766*** \\
               & {\scriptsize(0.0528)} & {\scriptsize(0.0536)} \\
$\eta_{6}$     & 0.0106      & 0.1813*** \\
               & {\scriptsize(0.0416)} & {\scriptsize(0.0429)} \\
$\eta_{7}$     & 0.1189**    & 0.4580*** \\
               & {\scriptsize(0.0510)} & {\scriptsize(0.0536)} \\
$\eta_{8}$     & 0.0027      & 0.0950*** \\
               & {\scriptsize(0.0236)} & {\scriptsize(0.0238)} \\
$\eta_{9}$     & -0.0515*    & -0.1444*** \\
               & {\scriptsize(0.0261)} & {\scriptsize(0.0271)} \\
$\eta_{10}$    & 0.0482**    & 0.1510*** \\
               & {\scriptsize(0.0213)} & {\scriptsize(0.0226)} \\
$\alpha_{1}$   & 5.3587      & 15.6740** \\
               & {\scriptsize(6.2250)} & {\scriptsize(6.4990)} \\
$\alpha_{2}$   & -3.7802     & -15.6690*** \\
               & {\scriptsize(3.5350)} & {\scriptsize(3.6969)} \\
\bottomrule
\end{tabular}
\end{threeparttable}
\vspace{-0.5cm}\\\bigskip\RaggedRight\singlespacing
\footnotesize Notes: Robust standard errors in parentheses. $^{***}p<0.01$, $^{**}p<0.05$, $^{*}p<0.1$.
\end{table}

Most notably, prior to the adjustment, neither the proxy nor most of the price interaction effects are significant. After accounting for price endogeneity due to the aggregated individual corner solutions, most effects are significant and, in the case of the proxy parameters, the magnitude is particularly expressive. To analyse the implications of the consideration set constraints for the adjusted model where it comes to price elasticities, an analysis of the resulting mark-ups is presented below. The revenue-weighted mean mark-ups are presented both for the case where goods are assumed to be priced independently (product-level pricing assumption) and by their respective supermarket (retailer-level pricing assumption). 

\begin{figure}[H]
    \centering
    \caption{Revenue-weighted mean mark-ups - product- and supermarket-level pricing}
    \begin{subfigure}[a]{.47\linewidth}
    \centering
    \includegraphics[width=\linewidth]{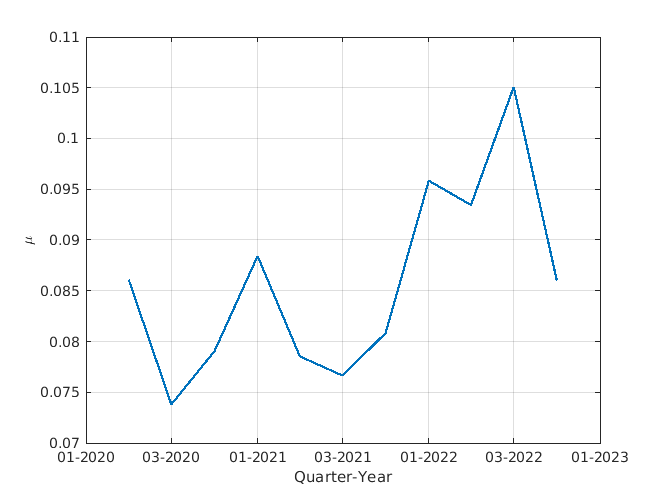}
    \end{subfigure}
    \begin{subfigure}[a]{.47\linewidth}
    \centering
    \includegraphics[width=\linewidth]{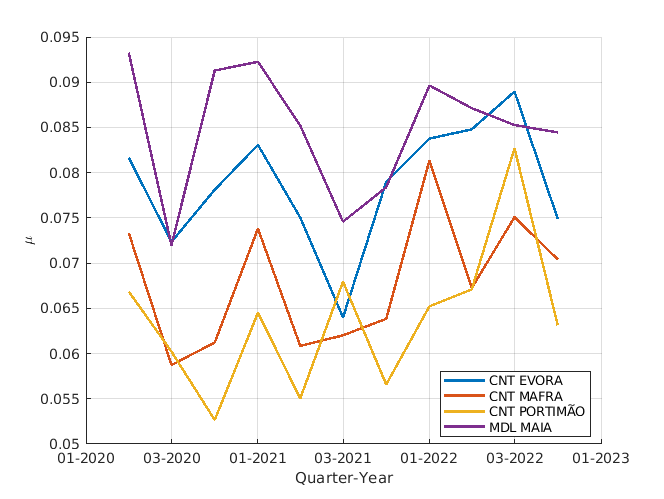}
    \end{subfigure}
\end{figure}

In each case the average mark-up shows some seasonality, with an upward trajectory before returning to baseline in the last quarter of the sample. On the left, throughout the sample period, mark-ups average around 8.5\% though under some volatility - perhaps less than one may othewise expect for this sample period. We can observe a distinct peak in the first quarter of 2021 ($\approx$8.5\%), and another peak in the third quarter of 2022 ($\approx$10.5\%). On the right, volatility tracks that on the left - Supermarket Maia averages a 8.5\% mark-up; Hypermarket Évora at 7.5\%; Hypermarket Mafra at 7\%, and Hypermarket Portimão at 6.5\%.\\

\begin{figure}[H]
\caption{Revenue-weighted four-quarter moving average mark-ups: U.S. survey estimate and sample}
\centering
\includegraphics[width=0.7\linewidth]{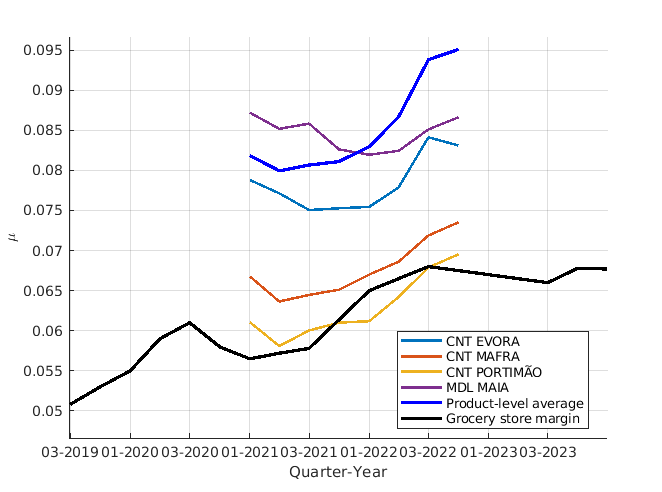}
\label{fig:moveavgPTUS}
\end{figure}

Regardless of pricing assumption, the results obtained in the present paper match survey estimates of United States grocery store margins remarkably well.\footnote{Data from U.S. Census Bureau's Quarterly Financial Report (QFR) for Retail Trade, reproduced from a report by the White House Council of Economic Advisors, originally obtained at: https://bidenwhitehouse.archives.gov/cea/written-materials/2024/06/20/update-grocery-price-inflation-has-cooled-substantially.} U.S. data is used here due to the unavailability of quarterly figures for Portugal.\footnote{The data which does exist is annual, see Banco de Portugal's Quadros do Setor in https://www.bportugal.pt/QS/qsweb/Dashboards. It roughly corroborates the U.S. trends.} Figure \ref{fig:moveavgPTUS} provides a side-by-side comparison of the four-quarter moving average mark-up in our sample and in the U.S. for our sample period.\\

The trends are near identical - the U.S. trend leading the Portuguese trend by about a quarter - though levels differ cross-country. U.S. mark-ups see a trough of 5.7\% in the first quarter of 2021 and a peak of 6.8\% in the third quarter of 2022: roughly half of those observed under product-level pricing, though it matches the lower bound of the supermarket mark-ups under retail-pricing. Any minor differences may nonetheless lie in grocery store margins excluding supplier mark-ups. Draganska, Klapper, and Villas-Boas (2010) find a rough 50/50 split in mark-ups between retailers and manufacturers in the German market for ground coffee, using scanner data from a national sample of stores belonging to six major retail chains. Portuguese and U.S. grocery store margins may also differ. Any differences are therefore within the expected interval for double marginalisation. Differences may also be explained by other costs accounted for in the grocery store profit margin, which are unobserved to researchers.\\

Revenue-weighted mark-up distributions are right-skewed, though there is some degree of heterogeneity across product categories. The following Figure provides a selection of these under the product-level pricing assumption:

\begin{figure}[H]
    \centering
    \caption{Average mark-up distribution - selected categories}
    \includegraphics[scale=0.6]{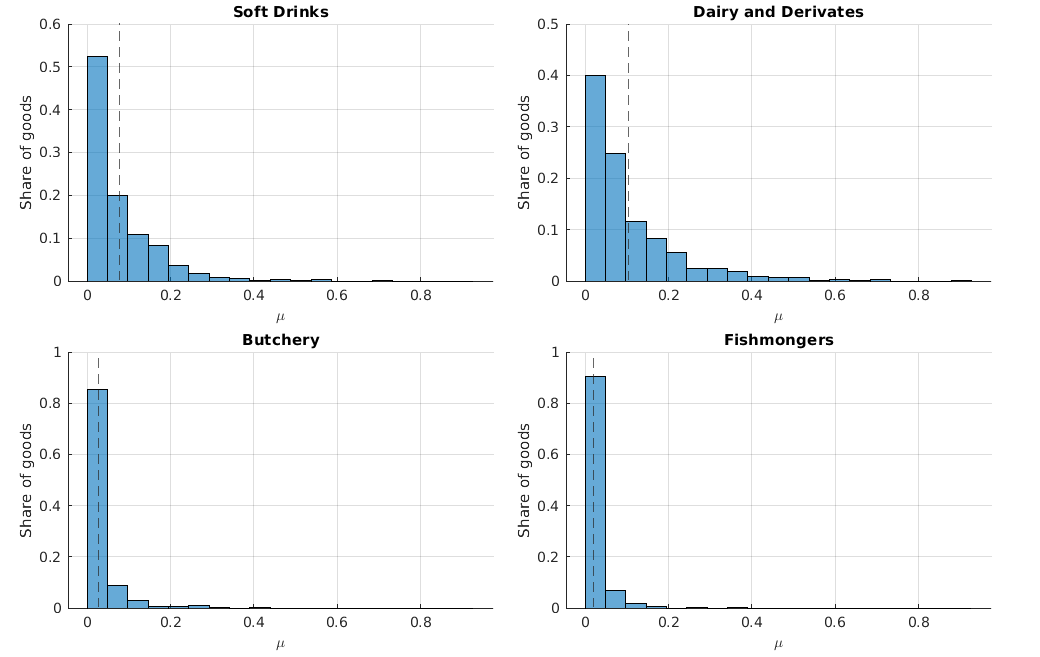}
    \label{meanpercmark-upsovertime}
\end{figure}

These results track some of the interaction effects revealed earlier. For example, "Butchery" ranks amongst the highest in own-category substitution exposure, while "Dairy and Derivates", despite ranking highly in own-category substitution exposure, also ranks well in own-category complementarity. "Fishmongers" rank low on own-category substitution and complementarity, but highly on cross-category substitution. "Soft Drinks" distribution is most likely connected to significant differentiation in the category, which includes "Beer", "Juice", "Water", and "Soda".\\

To obtain further insight on the distribution and significance of the mark-up trends, Figure \ref{meanpercmark-upsovertime} provides robust confidence intervals on the revenue-weighted mean mark-up estimates across mark-up percentiles, under the product-level pricing assumption.\\

\begin{figure}[H]
    \centering
    \caption{Revenue-weighted mean mark-ups - percentiles}
    \includegraphics[scale=0.6]{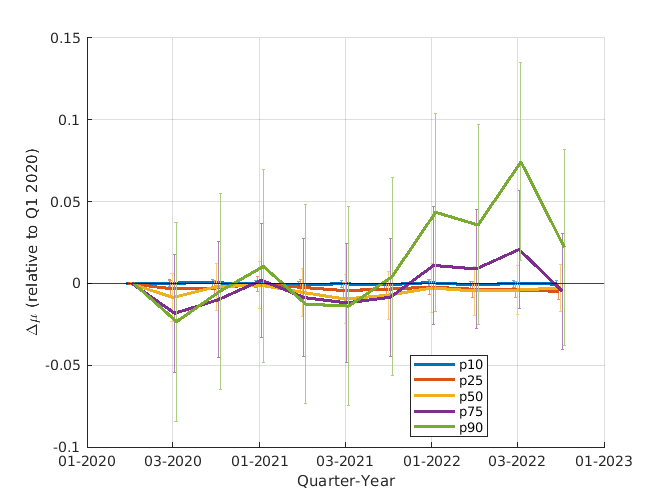}
    \label{meanpercmark-upsovertime}
\end{figure}

Mark-up changes remain insignificant for most goods across the sample period, except for a 7.4\% increase in the third quarter of 2022 amongst the highest percentile group, which seems to start to some extent, though non-significantly, two quarters prior (3.6\% and 4.3\% respectively), before dissipating the next period. This corresponds to the exact period where the CPI most closely aligns with sample inflation, likely due to supply chain disruption and an international energy shock. Notably, the increase in mark-ups observed in 2022 is not driven by supermarket heterogeneity, as can be seen in Figure \ref{supermarketreveweighmark-upsmmm} (here under the retailer-level pricing assumption):

\begin{figure}[H]
    \centering
    \caption{Revenue-weighted mean mark-ups - supermarkets}
    \includegraphics[scale=0.6]{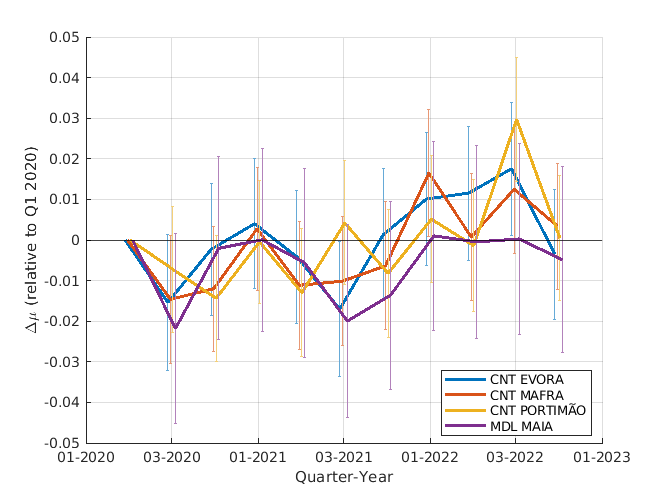}
    \label{supermarketreveweighmark-upsmmm}
\end{figure}

\begin{figure}[H]
    \centering
    \caption{Mark-up transition dynamics and transition shares over time}
    \includegraphics[scale=0.6]{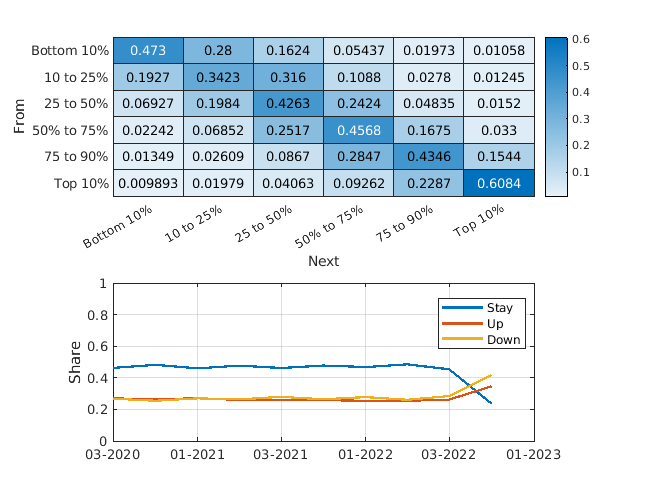}
    \label{supermarketreveweighmark-ups2}
\end{figure}

Across pricing specifications, the most robust trend shift throughout the sample period is the late mark-up raise around Q4 2022. To explore the causes of this apparent late jump in market power, I consider a few measures of product mobility across the mark-up distribution in Figure \ref{supermarketreveweighmark-ups2}. First, I recompute percentile bins within each category every quarter-year period, assign each product to a bin, and track transitions across adjacent quarter-years. A transition heatmap maps cross-time transitions, while a companion line plot shows the time path of said transitions.

Overall, we observe substantial time persistence, coupled with some degree of mean reversion. Products at the top end of the mark-up distribution, which we have noticed above to be a key driver of the overall mark-up increase observed at the end of our sample period, are especially persistent in their positioning. However, as can be seen in the line plot, persistence is shaken around the time where we have elsewhere observed a rise in market power, in mid-to-late 2022.\\

\begin{figure}[H]
    \centering
    \caption{Fixed-basket mark-up indices}
    \includegraphics[scale=0.6]{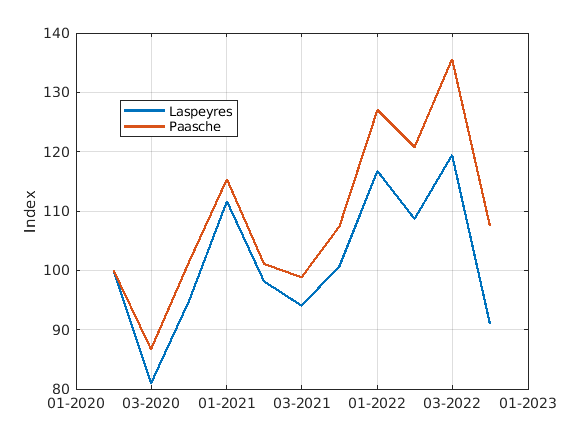}
    \label{supermarketreveweighmark-ups}
\end{figure}

To explore this further, in Figure \ref{supermarketreveweighmark-ups} we observe Laspeyres- and Paasche-type fixed-basket mark-up indices, based on initial and final revenue weights respectively.

Divergence between them diagnoses reallocation effects; and in fact, as of the third quarter of 2022, the Paasche index sat above the Laspeyres index, revealing shifting preferences. This reflects reports from industry players, who suggest two key factors as playing an important role in this change in behaviour: (i) brand switching as a result of stockouts; and (ii) a move away from in-person dining, which as of 2024 had yet to rebound.\footnote{See e.g. https://www.mckinsey.com/industries/consumer-packaged-goods/our-insights/state-of-consumer, and https://www.mckinsey.com/capabilities/growth-marketing-and-sales/our-insights/survey-us-consumer-sentiment-during-the-coronavirus-crisis.} A specific group of high-revenue products, specifically of a high mark-up percentile, appears to have decoupled from other goods in early 2022.\\

How do these results compare to the existing literature? Dopper, MacKay, Miller, and Stiebale (2024) estimates separate random coefficient mixed logit models by category and year for a panel dataset of consumer products in the US. In that paper, a mean of median category-level mark-ups - which overweights less revenue-relevant, lower-mark-up goods relative to the measure presented here, and would thus supposedly be lower than our figure - points to an average mark-up of 63\% in U.S. grocery store scanner data. Additional category-specific mark-up estimates from the literature are provided below:

\begin{table}[H]
\centering \footnotesize
\caption{Revenue-weighted median own-price elasticity by category}
\label{table:1}
\begin{tabular}{lccc}
\toprule
\textbf{Category} & \textbf{This Paper} & \textbf{Lit. Estimate} & \textbf{Citation}\\
\midrule
Beer & -44.93 & -4.74 & Miller and Weinberg (2017)\\
Breakfast cereal & -16.02 & -2.42 & Backus, Conlon, and Sinkinson (2021)\\
Yogurt & -6.99 & -4.05 & Hristakeva (2022) \\
\bottomrule
\end{tabular}
\vspace{-0.5cm}
\medskip\\\bigskip\RaggedRight\singlespacing
\footnotesize \textbf{Notes}: The Miller and Weinberg (2017) estimate is the median product-level elasticity obtained with the RCNL-1 specification. The Backus et al. (2021) estimate is the median product-level elasticity obtained with the “prices only” specification. Hristakeva (2022) reports a mean product-level elasticity from 2001–2010. My corresponding estimate is the revenue-weighted median own-price elasticity, across all periods, for the subcategories: "Soft Drinks" - "Beer"; "Breakfast" - "Cereal"; and "Dairy and Derivates" - "Yogurt and Desserts", net of "Desserts".
\end{table}

The relative level difference may be connected to a difference between Portuguese and American tastes; relative to Portugal, per capita the US consumes roughly 32.4\% more beer (Kirin Beer, 2020), roughly the same in breakfast cereal (Lopes et al., 2017; Lin et al., 2020), and 3-4 times less yogurt (Marramaque and Cardoso, 2020; Sebastian et al., 2025). The most noticeable feature of Table \ref{table:1} is however the magnitude difference between the measures in the literature and the findings in this paper. A source of the magnitude difference may in how the random coefficient mixed logit ignores cross-category effects. This has predictable implications for price elasticity estimates. Consider a stylised regression parameter for own-price effects, in the case where cross-category effects are ignored (or more broadly where market boundaries are understated):

\begin{equation}
    \hat{\beta}_i = \frac{Cov({q}_i,p_i)}{ Var(p_i)}=\frac{\sum_kJ_{ik}Cov(p_k,p_i)}{Var(p_i)}=J_{ii}+\sum_{k\neq i}J_{ik}\frac{Cov(p_k,p_i)}{Var(p_i)}
\end{equation}

for $J=\phi\cdot A(A'MA)^+A'$. $J_{ii}<0$ while $J_{ki}$ may be positive or negative depending on the product pair. Firms are strategic complements under Bertrand competition, so retailer prices co-move weakly positively across substitute goods and weakly negatively otherwise, suggesting $Cov({q}_i,p_i)\geq0$ and $Cov({q}_i,p_i)\leq0$ respectively. The cross-price terms $J_{ki}$ behave similarly, meaning that the summation contributes positively to $\hat \beta_i$. Thus, if cross-category price effects are ignored (or included, as in this paper), own-price effect estimates are biased towards (away from, respectively) zero; this underestimates (overestimates, respectively) price elasticities, possibly producing excessive (insufficient, respectively) estimates of market power and understating (overstating, respectively) pass-throughs. This also explains why the retailer-level mark-ups appear to be on average lower, as cross-price effects are directly incorporated into the optimal pricing. \\

Other arguments can help reconcile the rest of the literature with the reported margins in our sample and at U.S. grocery stores. Single-category research has in the past targeted categories where American suppliers' market power is especially evident, either due to market share concentration (e.g. soda, mark-ups estimated at between 19.9 and 63.9\%; Dubé, 2005) or brand proliferation by large multi-product firms (e.g. ready-to-eat cereals, mark-up estimated at 35.8\% under assumption of single-product firms; Nevo, 2001). In the case of Dopper et al. (2024), the sample ends in 2019 and observes stable prices, while our sample starts in early 2020 and contains significant price volatility. In any case, it may be to some extent hard to rationalise low grocery store profit margins, such as those observed in the US Census Bureau's Quarterly Financial Report, using mark-ups as high as those which have been previously reported in the literature.

\section{Discussion and Conclusion}

We have developed a novel way of thinking about consumer choice in differentiated-products markets where purchases commonly involve multiple goods, multiple units, and combinations across categories. The central idea is that consumers do not only choose quantities of individual goods from the full product space, but their consideration sets also incorporate combinations of goods. Modelling choice in this way makes it possible to accommodate realistic limits on attention within a tractable framework.\\

The main theoretical finding is that, even when basket choice itself is not uniquely pinned down, the induced demand over goods remains unique and linear. It can then be shown that consideration-set-constrained choice can provide a workable microfoundation for high-dimensional continuous demand systems. If consumers share proportional local demand slopes and the aggregate consideration set is sufficiently rich, then aggregate demand can be rationalised by the behaviour of a representative consumer within the same functional class, despite heterogeneity in preferences, price sensitivity, and consideration sets. This substantially relaxes the homogeneity requirements usually associated with representative-consumer results and helps address a long-standing concern in the linear-demand literature. It also provides the conceptual bridge to empirical work with market-level data, where the researcher observes aggregate outcomes but not the full structure of individual choice.\\

One remaining open question is the realism of assuming that consumers agree on what goods are (closer) substitutes or complementary. By focusing on common demand slopes, the theoretical model makes a testable claim. This clears the way for the use of non-anonymised transaction data to flexibly estimate individual price effects and test Eaton and Lipsey (1989)'s observation. This is left for future work.\\

Incidentally, the consideration set approach in this Chapter somewhat mirrors Lancaster (1966) and its treatment of "activities", the name given in that paper to shopping baskets. This paper can therefore also be understood as contributing to shedding light on this less well-known feature of that seminal paper: making said analysis mathematically tractable and applying it to further our understanding of consumer choice in more general settings.\\

We have also shown that aggregated linear demand can be estimated in a high-dimensional retail setting once the main empirical obstacles are addressed directly. I proposed an empirical strategy that combines a consideration-set-based proxy for substitution and complementarity effects with a 2SLS specification and a control-function correction for aggregated individual corner solutions. Once the Lagrangean constraints on shopping-basket choice are taken seriously, the resulting adjustments separate movements in demand driven by true price effects from those driven by changes in the set of active consumers. In doing so, the approach makes it possible to recover economically meaningful price elasticity parameters across a product assortment that would be difficult to study otherwise. The results suggest that empirical work with linear demand has been constrained less by the functional form itself than by the difficulty of estimating it credibly in large product spaces. Once dimensionality is reduced in a disciplined way, instruments are extended to cover rich price interactions, and aggregated corner solutions are explicitly incorporated, linear demand becomes a scalable and informative tool for studying competition in differentiated products markets. This is particularly valuable in settings where cross-category interactions, broad assortments, and evolving shopping baskets make conventional category-by-category approaches restrictive.\\

The empirical application of the paper finds that during a period of substantial demand and supply shocks to the Portuguese economy, mark-ups remained relatively stable. Post COVID, a significant jump occurs, though it is driven primarily by goods which were already high-mark-up prior to the increase. While these results are linked to changing tastes around that period, a more intense exploration of the drivers of these taste changes remains outside the scope of the paper. This leaves multiple unexplored possibilities for future research. One such possibility are stock-outs. Stock-outs were prevalent in this period. In the paper, I argue that they may have been one of the reasons why consumers drifted towards high-mark-up goods, raising them higher still. The story may not be so simple, as we may expect high-mark-up goods to have fewer substitutes, and therefore being expected to benefit less from shifting purchasing behaviour. It does however suggest that stock-outs may have a two-sided effect: relocating consumers to goods they value less than the ones they were buying, and doing so towards high-mark-up goods, whose mark-ups rise higher still as a result of said relocation and impose a further welfare loss. \\

Another interesting trend that goes under-discussed is the changing composition of the characteristics of goods in the supermarket. This has been reported anecdotally, but no systematic estimation of its effects has yet been made. The dataset available for this paper excluded data on specific product characteristics and inventory, meaning that this trend was not verifiable.\\

\normalsize 

\pagebreak

\section{References}
\small\singlespacing

Allen, R., and Rehbeck, J. (2022). Latent complementarity in bundles models. Journal of Econometrics, 228(2), 322-341.\\

Amir, R., Erickson, P., and Jin J. (2017). On the microeconomic foundations of linear demand for differentiated products. Journal of Economic Theory, 169, 641-665.\\

Armstrong, T. (2016). Large market asymptotics for differentiated product demand estimators with economic models of supply. Econometrica, 84(5), 1961-1980.\\

Atalay, E., Frost, E., Sorensen, A., Sullivan, C., and Zhu, W. (2023). Scalable demand and mark-ups (NBER Working Paper). National Bureau of Economic Research.\\

Backus, M., Conlon, C., and Sinkinson, M. (2021). Common Ownership and Competition in the Ready-to-Eat Cereal Industry. Working paper.\\

Bajari, P., Cen, Z., Chernozhukov, V., Manukonda, M., Vijaykumar, S., Wang, J., Huerta, R., Li, J., Leng, L., Monokroussos, G., and Wang, S. (2025). Hedonic prices and quality adjusted price indices powered by AI. Journal of Econometrics, 251, Article 106052.\\

Berry, S., Levinsohn, J., and Pakes, A. (1995). Automobile prices in market equilibrium. Econometrica, 63(4), 841-890.\\

Caplin, A., Dean, M., and Leahy, J. (2019). Rational inattention, optimal consideration sets, and stochastic choice. The Review of Economic Studies, 86(3), 1061–1094.\\

Caplin, A., Dean, M., and Martin, D. (2011). Search and Satisficing. American Economic Review, 101(7), 2011.\\

Chen Y., and Riordan, M. (2007). Price and variety in the spokes model. The Economic Journal, 117(522), 897-921.\\

Choné, P., and Linnemer, L. (2020). Linear demand systems for differentiated goods: Overview and user's guide. International Journal of Industrial Organization, 73, 1-18. \\

Christensen, L., Jorgenson, D., and Lau, L. (1975). Transcendental Logarithmic Utility Functions. American Economic Review, 65(3), 367-383.\\

Conlon, C. (2013). The Empirical Likelihood MPEC Approach to Demand Estimation. Working paper.\\

Crawford, G., Griffith, R., and Iaria, A. (2021). A survey of preference estimation with unobserved choice set heterogeneity," Journal of Econometrics, 222(1, A), 4-43.\\

Deaton, A., and Muellbauer, J. (1980). An almost ideal demand system. American Economic Review, 70, 312-326.\\

Dopper, H., MacKay, A., Miller, N., and Stiebale, J. (2024). Rising mark-ups and the Role of Consumer Preferences. Journal of Political Economy, 133(8).\\

Draganska, M., Klapper, D., and Villas-Boas, S. (2010). A Larger Slice or a Larger Pie? An Empirical Investigation of Bargaining Power in the Distribution Channel. Marketing Science, 9(1), 57-74.\\

Dubé, J. (2005). Product Differentiation and Mergers in the Carbonated Soft Drink Industry. Journal of Economics \& Management Strategy, 14(4), 879-904.\\

Eaton, B., and Lipsey, R. (1989). Product differentiation. In: Schmalensee, R.E., Willig, R.D. (Eds.), Handbook of industrial organization. North-Holland, Amsterdam, 723-768.\\

Eliaz, K., and Spiegler, R. (2011). Consideration Sets and Competitive Marketing. Review of Economic Studies, 78(1), 235-262.\\

Ershov, D., Laliberté, J.-W., Marcoux, M., and Orr, S. (2025). Estimating complementarity with large choice sets: An application to mergers. The RAND Journal of Economics, 56(4), 689–707.\\

Federgruen, A., and Hu, M. (2021). Technical note - Global robust stability in a general price and assortment competition model. Operations Research, 69(1), 164–174.\\

Fosgerau, M., Monardo, J., and de Palma, A. (2024). The Inverse Product Differentiation Logit Model. American Economic Journal: Microeconomics, 16(4), 329-370.\\

Fox, T., and Lazzati, N. (2017). A note on identification of discrete choice models for bundles and binary games. Quantitative Economics, 8(3), 1021-1036.\\

Gandhi, A., and Nevo, A. (2021). Empirical models of demand and supply in differentiated products industries. In K. Ho, A. Hortaçsu, and A. Lizzeri (Eds.), Handbook of Industrial Organization, 4(1), 63-139. Elsevier.\\

Gentzkow, M. (2007). Valuing new goods in a model with complementarity: Online newspapers. American Economic Review, 97(4), 713-744.\\

Gorman, W. (1953). Community preference fields. Econometrica, 21(1), 63–80.\\

Gorman, W. M. (1961). On a class of preference fields. Metroeconomica, 13(2), 53–56.\\

Hauser, J. (2014). Consideration-set Heuristics. Journal of Business Research, 67(8), 1688-1699.\\

Hauser, J. and Wernerfelt, B. (1990). An Evaluation Cost Model of Consideration Sets. The Journal of Consumer Research, 16(4), 393-408.\\

Hausman, J., Leonard, G., and Zona J. (1994). Competitive analysis with differentiated products. Annales d'Économie et de Statistique, 34, 159-180.\\

Hausman, J. (1996). Valuation of new goods under perfect and imperfect competition. In: The Economics of New Goods, 207-248. National Bureau of Economic Research, Inc.\\

Hildenbrand, W. (1983). On the "Law of Demand". Econometrica, 51(4), 997-1019.\\

Hoberg, G., and Phillips, G. (2016). Text-Based Network Industries and Endogenous Product Differentiation. Journal of Political Economy, 124(5), 1423-1465. \\

Hotelling, H. (1929). Stability in competition. The Economic Journal, 39(153), 41-57. \\

Hristakeva, S. (2022). Vertical Contracts with Endogenous Product Selection: An Empirical Analysis of Vendor Allowance Contracts. Journal of Political Economy, 130(12), 3202-3252.\\

Iaria, A., and Wang, A. (2019). Identification and Estimation of Demand for Bundles. Available at SSRN: https://ssrn.com/abstract=3458543 \\

Kirin Beer, (2020). Global Beer Consumption by Country in 2019. Tokyo: Kirin Holdings Company, Limited.\\

Koch, I. (2013). Analysis of multivariate and high-dimensional data. Cambridge University Press.\\

Lancaster, K. (1966). A new approach to consumer theory. Journal of Political Economy, 74, 132-157.\\

Lanier, J., Large, J., and Quah, J. (2023). Estimating very large demand systems. Available at https://ora.ox.ac.uk/objects/uuid:13accd61-5e96-46e8-849a-fdd7c627cbfd \\

Lee, Y. (2024). Shrinkflation: Evidence on Product Downsizing and Consumer Response. Kilts Center at Chicago Booth Marketing Data Center Paper, Available at SSRN: https://ssrn.com/abstract=5053745 or http://dx.doi.org/10.2139/ssrn.5053745.\\

Lee, L., and Pitt, M. (1986). Microeconometric Demand System with Binding Nonnegativity Constraints: The Dual Approach. Econometrica, 54(5), 1237-1242.\\

Lewbel, A., and Nesheim, L. (2019). Sparse demand systems: corners and complements. Boston College Working Papers in Economics 1005, Boston College Department of Economics.\\

Lin, B., Variyam, J., Allshouse, J., and Cromartie, J. (2003). Food and agricultural commodity consumption in the United States: Looking ahead to 2020 (Agricultural Economic Report No. 820). U.S. Department of Agriculture, Economic Research Service.\\

Lopes, C., Torres, D., Oliveira, A., Severo, M., Alarcão, V., Guiomar, S., Mota, J., Teixeira, P., Ramos, E., Rodrigues, S., Vilela, S., Oliveira, L., Nicola, P., Soares, S., Andersen. L (2017). Inquérito Alimentar Nacional e de Atividade Física (IAN-AF) 2015-2016. Universidade do Porto.\\

Magnolfi, L., McClure, J., and Sorensen, A. (2025). Triplet embeddings for demand estimation. American Economic Journal: Microeconomics, 17(1), 282-307.\\

Manchanda, P., Ansari, A., and Gupta, S. (1999). The "Shopping Basket": A Model for Multicategory Purchase Incidence Decisions. Marketing Science, 18(2), 95-114.\\

Manzini, P., Mariotti, M., and Ülkü, L. (2019). Stochastic Complementarity. The Economic Journal, 129(619), 1343-1363.\\

Marramaque, M., and Cardoso, F. (2021). Dairy Focus 1/2021: Portuguese EU Presidency. European Dairy Association.\\

Matějka, F., and MacKay, A. (2015). Rational Inattention to Discrete Choices: A New Foundation for the Multinomial Logit Model. American Economic Review, 105(1), 272-98.\\

McFadden, D. (1973) Conditional Logit Analysis of Qualitative Choice Behavior. In: Zarembka, P., Ed., Frontiers in Econometrics, Academic Press, 105-142.\\

Miller, N., and Weinberg, M. (2017). Understanding the Price Effects of the Miller-Coors Joint Venture. Econometrica, 85(6), 1763-1791.\\

Moore, E. (1920). On the reciprocal of the general algebraic matrix. Bulletin of the American Mathematical Society, 26(9), 394-395.\\

Nevo, A. (2001). Measuring market power in the ready-to-eat cereal industry. Econometrica, 69(2), 307-342.\\

Pellegrino, B. (2025). Product differentiation and oligopoly: A network approach. American Economic Review, 115(4), 1170-1225.\\

Penrose, R. (1955). A generalized inverse for matrices. Proceedings of the Cambridge Philosophical Society. 51(3), 406-413. \\

Pinkse, J., Slade, M. E., and Brett, C. (2002). Spatial price competition: A semiparametric approach. Econometrica, 70(3), 1111–1153.\\

Pinkse, J., and Slade, M. (2004). Mergers, brand competition, and the price of a pint. European Economic Review, 48(3), 617-643.\\

PORDATA. (2025). Contemporary Portugal Database. Retrieved from http://www.pordata.pt.\\

Salop, S. (1979). Monopolistic competition with outside goods. The Bell Journal of Economics, 10(1), 141-156. \\

Sebastian, R., Hoy, M., Murayi, T., Goldman, J., and Moshfegh, A. (2024,). Breakfast Consumption by U.S. Adults: What We Eat in America, NHANES 2017 – March 2020 (Dietary Data Brief No. 59). U.S. Department of Agriculture, Agricultural Research Service.\\

Simon, H. (1957). Models of Man: Social and Rational. New York: John Wiley and Sons.\\

Sovinsky, M. (2008). Limited Information and Advertising in the U.S. Personal Computer Industry. Econometrica, 76(5), 1017-1074.\\

Spence, M. (1976). Product Selection, Fixed Costs, and Monopolistic Competition. Review of Economic Studies, 43(2), 217-235.\\

Sun, T. (2024). Bundle choice model with endogenous regressors: An application to soda tax [arXiv preprint]. arXiv.\\

Theil, H. (1965). The Information Approach to Demand Analysis. Econometrica, 33(1), 67-87.\\

Thomassen, Ø., Smith, H., Seiler, S., and Schiraldi, P. (2017). Multi-category competition and market power: A model of supermarket pricing. American Economic Review, 107(8), 2308-51.\\

Tian, Y., Lautz, S., Wallis, A., and Lambiotte, R. (2021). Extracting complements and substitutes from sales data: A network perspective. EPJ Data Science, 10(45).\\

Ushchev, P., and Zenou, Y. (2018). Price competition in product variety networks. Games and Economic Behavior, 110, 226-247.\\

Vives, X. (1987). Small Income Effects: A Marshallian Theory of Consumer Surplus and Downward Sloping Demand. Review of Economic Studies, 54(1), 87-103.\\

Wales, T., and Woodland, A. (1983). Estimation of consumer demand systems with binding non-negativity constraints. Journal of Econometrics, 21(3), 263-285.\\

Wright P., and Barbour, F. (1977). Phased Decision Strategies: Sequels to an Initial Screening. Stanford School of Business Working Paper No. 353.\\

\pagebreak

\appendix

\section{Appendix}
\small\singlespacing

\subsection{Proofs}

\textit{Proposition 1.} The first part follows directly from linear dependence in the rows of a less-than-full-row-rank matrix, and helps us isolate the impact of the requirement that $\boldsymbol{q}_i$ be a linear function of $A_i$. The second part of the proposition can be proved as follows. Suppose $\mathrm{cone}(A_i) \subsetneq \mathbb{R}^K$. Then there exists at least one vector $\boldsymbol{x}_i \in \mathbb{R}^K$ such that $\boldsymbol{x}_i \notin \mathrm{cone}(A_i)$. In particular, consider the standard basis vectors $\boldsymbol{e}_1, \ldots, \boldsymbol{e}_K$. If all of these vectors were in $\mathrm{cone}(A_i)$, then by taking nonnegative linear combinations, we could generate any vector in $\mathbb{R}^K$, implying $\mathrm{cone}(A_i) = \mathbb{R}^K$, a contradiction; $\mathrm{cone}(A_i) \subsetneq \mathbb{R}^K$ implies that at least one $\boldsymbol{e}_k \notin \mathrm{cone}(A_i)$. Conversely, suppose there exists some $k$ such that $\boldsymbol{e}_k \notin \mathrm{cone}(A_i)$. Then, the cone cannot generate all vectors in $\mathbb{R}^K$, because it fails to generate the direction corresponding to good $k$ alone. Thus, $\mathrm{cone}(A_i)$ must be a strict subset: $\mathrm{cone}(A_i) \subsetneq \mathbb{R}^K$. \qedwhite\\

\textit{Lemma 1.} By construction, $\boldsymbol{x}'M\boldsymbol{x}>0$ for all non-zero $\boldsymbol{x}$. $\boldsymbol{y}'A'MA\boldsymbol{y}\geq0$ because, where $A\boldsymbol{y}\neq0$, $\boldsymbol{y}'A'MA\boldsymbol{y}=\boldsymbol{x}'M\boldsymbol{x}>0$ for non-zero $\boldsymbol{x}$. If $A\boldsymbol{y}=0$, $\boldsymbol{x}'M\boldsymbol{x}=0$ because $\boldsymbol{x}=A\boldsymbol{y}=0$. \qedwhite\\

\textit{Proposition 2.} Multiplying both sides by $A_i$, $\boldsymbol{q}_i=T_iA_i\boldsymbol{\hat{z}}_i$ and $A_i(I-(A_i'M_iA_i)^+(A_i'M_iA_i))\boldsymbol{y}=0$, as $I-(A_i'M_iA_i)^+(A_i'M_iA_i)$ is in the null space of $A_i'M_iA_i$, which itself matches the null space of $A_i$. $A_i$ and $A_i'MA_i$ share a null space as: if $A_i\boldsymbol{y}=0$, then $A'_iM_iA_i\boldsymbol{y}=0$ trivially; $\boldsymbol{y}'A_i'M_iA_i\boldsymbol{y}=(A_i\boldsymbol{y})'M_i(A_i\boldsymbol{y})=0$ if $A\boldsymbol{y}=0$. Since $M_i$ is assumed positive definite, the only way for us to have $(A_i\boldsymbol{y})'M_i(A_i\boldsymbol{y})=0$ is precisely if $A\boldsymbol{y}=0$. Last, $\xi$ is shown to not constrain $\boldsymbol{q}_i$ - i.e. $\xi=0$. This is because, for an individual consumer and shopping basket choice vector $\boldsymbol{\hat z}_i$, the unit-normalisation multiplier $\xi$ does not constrain $\boldsymbol{q}_i$ - i.e. $\xi=0$. We know this due to the following. From an individual consumer's optimisation in $T_i$, the stationarity condition is:

\begin{equation}
    \boldsymbol{\hat z}_i'A_i'(\boldsymbol{\delta}_i+\phi_i \boldsymbol{p})-T_i\boldsymbol{\hat z}_i'A'_iM_iA_i\boldsymbol{\hat z}_i=0
\end{equation}

From (\ref{z_FOC}), we have the stationarity condition from an individual consumer in $\boldsymbol{\hat z}_i$:

\begin{equation}
    A_i'(\boldsymbol{\delta}+\phi_i\boldsymbol{p})-T_i(A_i'M_iA_i)\hat{\boldsymbol{z}}_i+\frac{1}{T_i}(\boldsymbol{\hat\lambda_i}+\xi_i\boldsymbol{1})=0
\end{equation}

If we multiply the latter by $\boldsymbol{\hat z}_i'$, we can take advantage of (i) the complementary slackness condition such that $\boldsymbol{\hat z}_i'\boldsymbol{\hat \lambda}_i=0$, and (ii) the NORM constraint such that $\boldsymbol{1}'\boldsymbol{z}=1$, so:

\begin{equation}
    \boldsymbol{\hat z}_i'A_i'(\boldsymbol{\delta}+\phi_i\boldsymbol{p})-T_i\boldsymbol{\hat z}_i'(A_i'M_iA_i)\hat{\boldsymbol{z}}_i+\xi_i/T_i=0
\end{equation}

For both stationarity conditions to hold, we thus must have $\xi_i=0$. \\

Note that, should we have an additional condition restricting some elements of $\boldsymbol{\hat z}_i$ to $0$, the individual consumer's Lagrangean would include a zero-purchase multiplier, e.g. $\boldsymbol{\gamma}_i$. By construction, such a vector's elements would be positive only where $\boldsymbol{\hat z}_i$'s elements are equal to $0$ and vice-versa, so such a term would drop out once multiplied by $\boldsymbol{\hat z}_i'$. The finding that $\xi_i=0$ would therefore apply also under that more complex setting. \qedwhite\\

\textit{Lemma 2.} Regarding symmetry: for $M$ positive definite and symmetric (by construction), $A'MA$ is symmetric - $(A'MA)'= A'M'A = A'MA$. By eigenvalue decomposition, $A'MA=U\Lambda U'$, and $(A'MA)^+=U\Lambda^+U'$, where $\Lambda^+$ is formed by replacing each non-zero $\lambda,\forall i$ with $1/\lambda_i$, and leaving zeros intact. Since $\Lambda^+$ is diagonal, it is symmetric, and therefore $(A'MA)^+$ is also symmetric. Lastly, if $(A'MA)^+$ is symmetric, $A(A'MA)^+A'$ is also symmetric - $(A'(A'MA)^+A')'= A'((A'MA)^+)'A' = A(A'MA)^+A'$. Regarding positive semi-definiteness: If $A'MA$ is positive semi-definite, $(A'MA)^+$ is also positive semi-definite. As shown above, $A'MA$ is symmetric and can be diagonalised via eigenvalue decomposition, and the pseudoinverse inverts the non-zero eigenvalues while leaving the zero eigenvalues at zero. The inversion does not change the signs of the eigenvalues, so $\forall \lambda\geq0$ still holds. If $(A'MA)^+$ is positive semi-definite, $\boldsymbol{x}'(A'MA)^+\boldsymbol{x}\geq0$. This is the case for $\boldsymbol{x}=A'\boldsymbol{y}\neq0$. For $\boldsymbol{x}=A'\boldsymbol{y}=0$, $\boldsymbol{x}'(A'MA)^+\boldsymbol{x}=0$. Therefore, $\boldsymbol{x}'(A'MA)^+\boldsymbol{x}'=\boldsymbol{y}'A(A'MA)^+A'\boldsymbol{y}\geq0$ for all non-zero (and zero) $\boldsymbol{y}$. \qedwhite\\

\textit{Proposition 3.} An aggregate consideration set $A$ allows us to re-write the Lagrangian of each consumer $i$ in terms of $\boldsymbol{z}_i$:

\begin{equation}
\begin{aligned}
    \mathcal{L}(\boldsymbol{z}_i)=& T_i\boldsymbol{z}_i'A'\boldsymbol{\delta}_i  - \frac{1}{2} T_i^2\boldsymbol{z}_i' (A' M A) \boldsymbol{z}_i
- \phi_i \big( Y_i - T_i \boldsymbol{p}' A \boldsymbol{z}_i\big)\\
    &+ \boldsymbol{\lambda}_i' \boldsymbol{z}_i+\boldsymbol{\gamma}_i'\boldsymbol{z}_i
\end{aligned}
\end{equation}

Note that the redimensioning and inclusion of a zero-purchase multiplier is addressed in the proof of \textbf{Proposition 2}: $\xi_i=0$ here too. The individual KKT conditions for $\boldsymbol{z}_i$ will then look something like this:

\begin{equation}\label{KKTInd}
    A'\boldsymbol{\delta}_i-T_i(A'MA)\boldsymbol{z}_i+\phi_iA'\boldsymbol{p}+\frac{1}{T_i}(\boldsymbol{\lambda}_i+\boldsymbol{\gamma}_i)=0\quad\boldsymbol{\lambda}_i\geq0\quad\boldsymbol{{z}}_i\geq0\qquad \boldsymbol{\lambda}\circ\boldsymbol{{z}}_i=\boldsymbol{0}
\end{equation}

Stacking and collapsing the system of individual stationarity conditions, we get an aggregate:

\begin{equation}\label{stackandcoll}
    A'(\sum_{i=1}^N\boldsymbol{\delta}_i)-(A'MA)(\sum_iT_i\boldsymbol{z}_i)+A'(\sum_{i=1}^N\phi_i)\boldsymbol{p}+\sum_{i=1}^N\frac{1}{T_i}(\boldsymbol{\lambda}_i+\boldsymbol{\gamma}_i)=0
\end{equation}

However, this aggregate cannot be rationalised as the stationarity condition for a utility-maximising representative consumer, because the complementary slackness condition is not satisfied in general:

\begin{equation}\label{compslackcond}
    (\sum_{i=1}^N \boldsymbol{z}_i)\circ(\sum_{i=1}^N\boldsymbol{\lambda}_i)\neq0
\end{equation}

This is because, in general, we have both $\sum_{i=1}^N \boldsymbol{z}_i>0$ and $\sum_{i=1}^N\boldsymbol{\lambda}_i>0$ in the population: at least one shopping basket is bought \textit{and} not bought at least once. Yet for an identical stationarity condition produced by a representative consumer in $\boldsymbol{z}$ and $\boldsymbol{\Lambda}$, \textbf{Proposition 2} says the KKT conditions are satisfied. This is the general result known in the academic literature.\\

Our setup is slightly different however, incorporating consideration sets. Could we use these to find a way to conciliate this contradiction? Let us go step by step.\\

First, because of our condition that $\boldsymbol{q}_i=T_iA_i\boldsymbol{z}_i,\ \forall i$ (and therefore $\boldsymbol{Q}=TA\boldsymbol{z}$, for $\boldsymbol{z}=\frac{1}{T}\sum_{i=1}^NT_i\boldsymbol{z}_i$), there are infinitely many $\boldsymbol{z}$ one can have where this condition is satisfied: e.g. any $\boldsymbol{y}$ such that $\boldsymbol{Q}=TA(\boldsymbol{z}+\boldsymbol{y})$ and $ A\boldsymbol{y}=0$ would do.\\

Second, note that, beyond the above condition on $\boldsymbol{z}$, for the purposes of understanding how shopping basket choice affects aggregate goods consumption $\boldsymbol{Q}$, it is not the stationarity condition in ($\ref{stackandcoll}$) that binds, but rather that which we obtain after we multiply both sides by $A(A'MA)^+$ to obtain the aggregate demand function:

\begin{equation}
    \boldsymbol{Q}=A(A'MA)^+[A'(\sum_i\boldsymbol{\delta}_i+\phi_i \boldsymbol{p})+\sum_i\frac{1}{T_i}(\boldsymbol{\lambda}_i+\boldsymbol{\gamma}_i)]
\end{equation}

The $\boldsymbol{\Lambda}$ need not be in $\mathrm{col}(A')$. The same applies to a representative consumer's individual demand function, which we can write as:

\begin{equation}
\begin{aligned}
    \boldsymbol{q}^{rep}=&A(A'MA)^+[A'(\sum_i\boldsymbol{\delta}_i+\phi \boldsymbol{p})+\boldsymbol{\Lambda}+\boldsymbol{\Gamma}]\\
\end{aligned}
\end{equation}

Our goal is therefore to determine, for $\phi=\sum_{i=1}^N \phi_i$, whether there are any vectors $\boldsymbol{\Lambda}$ and $\boldsymbol{\Gamma}$ for which

\begin{equation}
    A(A'MA)^+[\sum_i\frac{1}{T_i}(\boldsymbol{\lambda}_i+\boldsymbol{\gamma}_i)] = A(A'MA)^+[\boldsymbol{\Lambda}+\boldsymbol{\Gamma}]
\end{equation}

holds. Turns out, yes. To see this, note that the $\boldsymbol{\Lambda}$ and $\boldsymbol{\Gamma}$ are not separately identified. For infinitely many $\boldsymbol{z}$, say $\boldsymbol{z}^\#$, we can then set $\boldsymbol{\Lambda}=\boldsymbol{\Lambda}^\#$ and $\boldsymbol{\Gamma}=\boldsymbol{\Gamma}^\#$ as follows:

\begin{equation}
    \Lambda_j^\#=\begin{cases}
        0 & \text{if }z_j^\#>0\\
        \sum_i\frac{1}{T_i}({\lambda}_{ij}+{\gamma}_{ij})& \text{if }z_j^\#=0\text{ and }\sum_i\frac{1}{T_i}({\lambda}_{ij}+{\gamma}_{ij})\geq0\\
        0 & \text{if }z_j^\#=0\text{ and }\sum_i\frac{1}{T_i}({\lambda}_{ij}+{\gamma}_{ij})< 0\\
    \end{cases}, \ \forall j=1,\ldots,J
\end{equation}
\begin{equation}
    \boldsymbol{\Gamma}^\#= \sum_i\frac{1}{T_i}(\boldsymbol{\lambda}_{i}+\boldsymbol{\gamma}_{i})-\boldsymbol{\Lambda}^\#
\end{equation}

These generate the same $\boldsymbol{Q}$ as $\sum_{i=1}^N\frac{1}{T_i}\boldsymbol{\lambda}_i$ and $\sum_{i=1}^N\frac{1}{T_i}\boldsymbol{\gamma}_i$. It is trivial to see that $\boldsymbol{\Lambda}^\#$ obtained in such a way satisfy $\boldsymbol{\Lambda}^\#\circ \boldsymbol{z}^\#=0$ and $\boldsymbol{\Lambda}^\#\geq0$. Crucially, $\boldsymbol{\Gamma}^\#$ is what lets you keep the multiplier sum fixed while reallocating the nonnegative part into $\boldsymbol{\Lambda}^\#$ so that complementary slackness holds for the new values. This result applies locally - price changes imply a new set of multipliers to correct for. \qedwhite\\

\textit{Lemma 3.} If $A$ is full row rank, it can be shown that $A(A'MA)^+A'=M^{-1}$. Its singular value decomposition could be written as $A=U\Sigma V'$, for $U$ a $K\times K$ orthogonal matrix; $\Sigma=[\Sigma_r 0]$ a $K\times J$ matrix, with $\Sigma_r$ a $K\times K$ diagonal matrix containing the nonzero singular values of $A$, and $0$ a $K\times (J-K)$ zero block; and $V=[V_1 V_2]$ is a $J\times J$ orthogonal matrix where $V_1$ is a $J\times K$ matrix whose columns form the orthonormal basis for the row space of $A$, while $V_2$ is a $J\times (J-K)$ matrix spanning the null space of $A$. $A$ can then be re-written as $A=U\Sigma_r V_1'$. From here, $A(A'MA)^+A'=U\Sigma_rV_1'(V_1\Sigma_r U'MU\Sigma_rV_1')^+V_1\Sigma_r U'=U\Sigma_rV_1'V_1(\Sigma_r U'MU\Sigma_r)^{-1}V_1'V_1\Sigma_r U'=U\Sigma_r(\Sigma_r U'MU\Sigma_r)^{-1}\Sigma_r U'=U(U'MU)^{-1} U'=M^{-1}$ by the properties of orthogonal and diagonal matrices. \qedwhite\\

\textit{Lemma 4.} The sum of a positive definite with a positive semi-definite matrix is not always invertible. By theorem, however, the sum of a positive definite $\Omega$ with a positive semi-definite matrix $A'(A'MA)^+A'$, \textit{where the positive definite matrix is diagonal with strictly positive diagonal elements}, is itself positive definite, and therefore invertible:
$\boldsymbol{x}'\Omega+A(A'MA)^+A'\boldsymbol{x}=\boldsymbol{x}'\Omega\boldsymbol{x}+\boldsymbol{x}'A(A'MA)^+A'\boldsymbol{x}=\text{positive}+\text{non-negative}>0$. We can show that $\Omega$ is strictly positive in its diagonal. Express the diagonal matrix whose diagonal matches that of $A(A'MA)^+A'$ as $\Omega$. To guarantee strictly positive elements in the diagonal of $\Omega$, two requirements must be satisfied. Firstly, $\boldsymbol{v}_i\neq 0,\forall i$, for $\boldsymbol{v}_i=A'\boldsymbol{e}_i$, where $\boldsymbol{e}_i$ is the standard basis vector, and $\Omega_{ii} = \boldsymbol{e}_i'A(A'MA)^+A'\boldsymbol{e}_i = \boldsymbol{v}_i'(A'MA)^+\boldsymbol{v}_i$, where $\boldsymbol{e}_i$ "picks out" the $i$-th diagonal of $\Omega$. This is the case as long as row $i$ of $A$ has at least one non-zero term, which is true of every row in $A$ by definition, as otherwise said row would describe a product we have not observed in the data. Secondly, $\boldsymbol{v}_i'(A'MA)^+\boldsymbol{v}_i>0,\forall i$. Now, as we mentioned earlier, $(A'MA)^+$ is positive semi-definite. However, $\boldsymbol{v}_i$ is restricted to the column space of $A$ by construction, the same column space of $(A'MA)^+$. On the subspace where $A'MA$ is "active" (its column space), the pseudoinverse $(A'MA)^+$ behaves like a true inverse and is positive definite on that subspace. This ensures that whenever is $\boldsymbol{v}_i$ is nonzero (i.e., when it lies in the column space of $A$), we have $\boldsymbol{v}_i'(A'MA)^+\boldsymbol{v}_i>0,\forall i$. Given our proof of strictly positive diagonal elements in $\Omega$, $\Omega+A(A'MA)^+A'$ is invertible. \qedwhite\\

\pagebreak

\subsection{Bertrand-Nash equilibrium analysis for less-than-full row $A$}

Having built-up an understanding of how linear aggregate demand operates, let us consider the implications of a less-than-full-rank consideration set for the outcomes of (Bertrand) price competition. We will account for two extreme market structures: that (i) of a finite number of single-product firms, and (ii) of a single multi-product monopoly. Both settings will be shown to deliver well-behaved demand with a unique equilibrium.\\

Assume first that all $K$ differentiated products in the assortment are sold by $K$ different single-product firms. For simplicity of exposition, marginal costs are set to $c_i=0,\forall i\in \mathcal{K}$. The representative consumer takes prices as given. All results are shown assuming interior solutions only for simplicity.\\

\textit{\textbf{Proposition}: When each of the $K$ differentiated products is sold by a separate firm engaging in Bertrand (Nash) competition, the unique equilibrium price vector is}:

\begin{equation}\label{optpricesing}
    \boldsymbol{p^*} = -\frac{1}{\phi}[\Omega + A(A'MA)^+A' ]^{-1} A(A'MA)^+A' \boldsymbol{\delta}
\end{equation}

\textit{for} $\Omega=\text{diag}(\boldsymbol{a_1}'(A'MA)^+ \boldsymbol{a_1},\ \ldots,\ \boldsymbol{a_K}'(A'MA)^+ \boldsymbol{a_K})$.\hfill$\blacksquare$\\

\textit{Proof}: Consider a given firm $i$'s objective function:

\begin{equation}
    \pi_i(\boldsymbol{p})=p_iq_i(\boldsymbol{p}) =p_i\big[a_{i}'(A'MA)^+A'(\boldsymbol{\delta}+\phi \boldsymbol{p})\big]
\end{equation}

The set of first-order conditions for this problem across all firms may therefore be defined as 

\begin{equation}
    0 = A(A'MA)^+A'\boldsymbol{\delta} + 2\phi\Omega \boldsymbol{p} + \phi(A(A'MA)^+A'-\Omega) \boldsymbol{\hat{p}}
\end{equation}

To find the Bertrand (Nash) equilibrium, we look for the fixed point $\boldsymbol{p^*}$ such that $\boldsymbol{p} = \boldsymbol{\hat{p}} = \boldsymbol{p^*}$ - i.e. that point which defines the best response as a function of all competing firms' strategies. Uniqueness is enabled by the following property:\\

\textit{\textbf{Lemma}: $\Omega + A(A'MA)^+A' $ is invertible.}\hfill $\blacksquare$\\

\textit{Proof}: The sum of a positive definite with a positive semi-definite matrix is not always invertible. By theorem, however, the sum of a positive definite $\Omega$ with a positive semi-definite matrix $A'(A'MA)^+A'$, \textit{where the positive definite matrix is diagonal with strictly positive diagonal elements}, is itself positive definite, and therefore invertible:
$\boldsymbol{x}'\Omega+A(A'MA)^+A'\boldsymbol{x}=\boldsymbol{x}'\Omega\boldsymbol{x}+\boldsymbol{x}'A(A'MA)^+A'\boldsymbol{x}=\text{positive}+\text{non-negative}>0$. We can show that $\Omega$ is strictly positive in its diagonal. Express the diagonal matrix whose diagonal matches that of $A(A'MA)^+A'$ as $\Omega$. To guarantee strictly positive elements in the diagonal of $\Omega$, two requirements must be satisfied. Firstly, $\boldsymbol{v}_i\neq 0,\forall i$, for $\boldsymbol{v}_i=A'\boldsymbol{e}_i$, where $\boldsymbol{e}_i$ is the standard basis vector, and $\Omega_{ii} = \boldsymbol{e}_i'A(A'MA)^+A'\boldsymbol{e}_i = \boldsymbol{v}_i'(A'MA)^+\boldsymbol{v}_i$, where $\boldsymbol{e}_i$ "picks out" the $i$-th diagonal of $\Omega$. This is the case as long as row $i$ of $A$ has at least one non-zero term, which is true of every row in $A$ by definition, as otherwise said row would describe a product we have not observed in the data. Secondly, $\boldsymbol{v}_i'(A'MA)^+\boldsymbol{v}_i>0,\forall i$. Now, as we mentioned earlier, $(A'MA)^+$ is positive semi-definite. However, $\boldsymbol{v}_i$ is restricted to the column space of $A$ by construction, the same column space of $(A'MA)^+$. On the subspace where $A'MA$ is "active" (its column space), the pseudoinverse $(A'MA)^+$ behaves like a true inverse and is positive definite on that subspace. This ensures that whenever is $\boldsymbol{v}_i$ is nonzero (i.e., when it lies in the column space of $A$), we have $\boldsymbol{v}_i'(A'MA)^+\boldsymbol{v}_i>0,\forall i$. Given our proof of strictly positive diagonal elements in $\Omega$, $\Omega+A(A'MA)^+A'$ is invertible. \qedwhite\\

It then follows that:\\

\textit{\textbf{Proposition}: In the single-product firms setting, the unique optimal demand $\boldsymbol{Q}(\boldsymbol{p}^*)$, is defined by the expression:}

\begin{equation}\label{optquantsing}
    \boldsymbol{Q}(\boldsymbol{p}^*)=\Omega(\Omega+A(A'MA)^+A')^{-1}A(A'MA)^+A'\boldsymbol{\delta}
\end{equation}\hfill$\blacksquare$

\textit{Proof}: Result follows from substituting $\boldsymbol{p}^*$ onto $\boldsymbol{Q}(\boldsymbol{p})=A(A'MA)^+(\boldsymbol{\delta}+\phi \boldsymbol{p})$.\qedwhite\\

What if all $K$ differentiated products in the assortment are sold by a single multi-product monopolist retailer? \\

\textit{\textbf{Proposition}: When each of the $K$ differentiated products is sold by a single, multi-product, price-setting monopolist, a multiplicity of price equilibria arises:}

\begin{equation}\label{optpricemono}
    \boldsymbol{p^*} = -\frac{1}{2\phi}\boldsymbol{\delta}+(I-W^+W)\boldsymbol{y}
\end{equation}

\textit{with}

\begin{equation}
    W=\Omega + A(A'MA)^+A'+A(A'MA)^+A' \circ G=2A(A'MA)^+A'
\end{equation} 

\textit{for $\boldsymbol{y}\in \mathbb{R}^K$ and $G$ is a symmetric hollow $K\times K$ ownership matrix where each element $G_{ab}= 1$ if $a,b\in S$ (i.e. owned by the same firm), $0$ otherwise. In this case, $G_{monopolist}=\boldsymbol{1}\boldsymbol{1}'-I$.}\hfill $\blacksquare$\\

\textit{Proof}: A multi-product firm determines a vector of prices for its goods by maximising the following profit function:

\begin{equation}
    \pi_S(\boldsymbol{p})=\sum_{i\in S}p_i[a_{i}'(A'MA)^+A'(\boldsymbol{\delta}+\phi \boldsymbol{p})]
\end{equation}

for $S$ the subset of products $i=1,\ldots, K$ sold by said firm. The FOC can be expressed in matrix form as:
\vspace{-0.3cm}

\begin{equation}\footnotesize
    0=A(A'MA)^+A'\boldsymbol{\delta}+\phi(2\Omega + 2(G \circ A(A'MA)^+A'))\boldsymbol{p}+ \phi((A(A'MA)^+A'-\Omega)-G \circ A(A'MA)^+A')\hat{\boldsymbol{p}}
\end{equation}

To find the Bertrand (Nash) equilibrium, we operate in a manner identical to that pursued to find the equilibrium price in the single-product firms case. The $(I-W^+W)\boldsymbol{y}$ which arises as a result of the pseudo-inverse implies a multiplicity of mappings from initial marginal product utility to equilibrium prices. \qedwhite\\

The inclusion of an ownership matrix is problematic for the identification of a unique price equilibrium in more general market structures, as we cannot say anything \textit{ex-ante} about the definiteness of $G$. In fact, without additional assumptions, no closed-form solutions for multi-product oligopoly case exist for this model (more on this elsewhere in the Appendix). However, for the monopoly case in particular, the multiplicity of price equilibria are not just payoff-equivalent, but \textit{equilibrium-quantity equivalent}:\\

\textit{\textbf{Proposition}: In the monopoly setting, the unique optimal demand $\boldsymbol{Q}(\boldsymbol{p}^*)$, is defined by the expression:}

\begin{equation}\label{optquantmono}
    \boldsymbol{Q}(\boldsymbol{p}^*)=\frac{1}{2}A(A'MA)^+A'\boldsymbol{\delta}
\end{equation}

\textit{while the aggregate profit and consumer surplus take on the following unique closed-form expressions:}

\begin{equation}
    \begin{aligned}
        \Pi=\sum_i^N\pi_i &= \boldsymbol{Q}'(\boldsymbol{p}^*)\boldsymbol{p}^*=-\frac{1}{4\phi}\boldsymbol{\delta}'A(A'MA)^+A'\boldsymbol{\delta}
    \end{aligned}
\end{equation}

\begin{equation}
\begin{aligned}
    CS = \boldsymbol{Q}'(\boldsymbol{p}^*)\boldsymbol{\delta}-\frac{1}{2}\boldsymbol{Q}'(\boldsymbol{p}^*)M\boldsymbol{Q}(\boldsymbol{p}^*)+\phi \Pi = \frac{1}{8}\boldsymbol{\delta}'A(A'MA)^+A'\boldsymbol{\delta}
\end{aligned}
\end{equation}\hfill $\blacksquare$

\textit{Proof}: The proof is trivial once we substitute $\boldsymbol{p}^*$ onto $\boldsymbol{Q}(\boldsymbol{p})=A(A'MA)^+A'(\boldsymbol{\delta}+\phi\boldsymbol{p})$ as well as onto the representative consumer's utility function and the monopolist's aggregate profit function.\qedwhite 

\pagebreak

\subsection{Comparative statics}

Let us consider how consideration sets impact aggregate demand. By \textbf{Proposition 1}, we can separately address vector space restrictions (i.e., the rank of $A$) from the the broader condition which includes choice vector non-negativity. We do so by varying the row rank of $A$ and relying on \textbf{Lemma 4}. When $A$ is less-than-full row rank (i.e. when choice is constrained by the representative consumer's consideration set restriction on the consumption vector space):

\begin{equation}
\begin{aligned}
    \boldsymbol{Q}&=A(A'MA)^+ A'(\boldsymbol{\delta}+ \phi \boldsymbol{p})\ \ s.t.\ \ \boldsymbol{z}\geq0
        \ \ \\
        &=A(A'MA)^+ A'(\boldsymbol{\delta}+ \phi \boldsymbol{p})+A(A'MA)^+ \boldsymbol{\Lambda}(\boldsymbol{\delta}, \boldsymbol{p})\\
\end{aligned}
\end{equation}

When $A$ is full row rank, $span(A)=\mathbb{R}_{+}^K$, $A(A'MA)^+A'=M^{-1}$, and

\begin{equation}\label{NNportionbinds}
\begin{aligned}
    \boldsymbol{Q}&=M^{-1}(\boldsymbol{\delta}+ \phi \boldsymbol{p})\ \ s.t.\ \ \boldsymbol{z}\geq0
        \ \ \\
        &=M^{-1}(\boldsymbol{\delta}+ \phi \boldsymbol{p})+A(A'MA)^+\boldsymbol{\Lambda}(\boldsymbol{\delta}, \boldsymbol{p})\\
\end{aligned}
\end{equation}

The key difference between the equations above is that the unconstrained Slutsky matrix either depends on $A(A'MA)^+A'$ or $M^{-1}$. Can we say anything about how they differ structurally? Yes:\\

\textit{\textbf{Proposition}: Where $A$ is less-than-full row rank, relative to $M^{-1}$, $A(A'MA)^+A'$ has (weakly) lower-valued diagonal elements.}\hfill $\blacksquare$\\

\textit{Proof}: The trick for this proof is to use congruence. Note that $M^{-1}=M^{-1/2}IM^{-1/2}$ and $A(A'MA)^+A'=M^{-1/2}PM^{-1/2}$, for $P=\tilde{A}(\tilde{A}'\tilde{A})^+\tilde{A}'$ where $\tilde{A} = M^{1/2}A$ and $P$ is an orthogonal projection onto the column space of $\tilde{A}$. By the aforementioned properties of the $M$ matrix, this means $M^{-1}$ and $I$ are congruent, and so are $A(A'MA)^+A'$ and $P$. For our purposes, this allows us to compare $P$ and $I$ knowing that said comparison extends to their congruents. For example, consider the diagonal elements of $P$. The definiteness properties of $I$ and $P$'s congruents extend to them: $\boldsymbol{x}'P\boldsymbol{x}\leq\boldsymbol{x}'I\boldsymbol{x},\forall \boldsymbol{x}$. Since $\boldsymbol{x}'I\boldsymbol{x}=\boldsymbol{x}'\boldsymbol{x}=||\boldsymbol{x}||^2$, $P_{ii}=\boldsymbol{e}_i'P\boldsymbol{e}_i\leq\boldsymbol{e}_i'\boldsymbol{e}_i=||\boldsymbol{e}_i||^2=1$. As all diagonal elements of $I$ are equal to 1, this means all diagonal elements of $P$ are weakly smaller than $I$'s, and all diagonal elements of $A(A'MA)^+A'$ are weakly smaller than $M^{-1}$'s (one need only think of $\boldsymbol{x}=M^{-1/2}\boldsymbol{e}_i,\forall i$). Each individual diagonal element of $A(A'MA)^+A'$ is weakly smaller than the diagonal elements of $M^{-1}$, so the average of these is also smaller. \qedwhite\\

This is an \textit{attenuation effect}. Where consumers consider and choose over combinations of goods, individual goods and their characteristics drive their own demand (weakly) less, as that will depend on the wider set being purchased. In these cases, own-price makes (weakly) less of an impact than it would otherwise. This result can be extended also to the non-negativity condition, which imposes an attenuation effect of its own. \\

\textit{\textbf{Corollary:} Where the non-negativity condition binds, own-price effects are (weakly) smaller in magnitude.}\hfill $\blacksquare$\\

\textit{Proof}: The NN condition introduces an active set of baskets with $z_j>0$ and separates it from an inactive set where $z_j=0$. This effectively replaces the column space of $A$ with a smaller (active) subspace in the column space of $A_{z_j>0}\subseteq A$. As we saw, orthogonal projection onto a smaller subspace weakens the length of any component's projection, thus leading to diagonal entries (weakly) smaller in magnitude.\qedwhite\\

Important questions remain. No stronger claim can be made regarding the exact change in the characteristics or magnitude of the off-diagonal elements of $\phi A(A'MA)^+A'$ relative to $\phi M^{-1}$. Similarly, not much can be said of the impact of the different characteristics of the model on its equilibrium outcomes. To address these limitations, I run a Monte Carlo simulation. This allows us to systematically generate random environments, solve for equilibrium quantities and prices under different market structures, and evaluate different consideration set structures. While not constituting solid proof, these outcomes provide some degree of confidence as to the direction of some otherwise hard-to-prove effects, and constitute hypotheses for further research. The simulation is run over 1000 draws, and is divided into three steps: primitive sampling, Hessian matrix formation, and equilibrium computations.\\

In the first step, the representative consumer's consideration set $A$ is produced from a baseline of 10 goods and 51 baskets (roughly 5\% of all feasible combinations). After correcting for zero- and non-unique columns and rows via non-negative linear combinations of the remaining, up to 10 columns and rows are also added in such a way. The average draw has 14.8 goods and 55.5 baskets, for an average rank of 10 and an average 14.3 missing standard basis vectors. The marginal-utility-of-income coefficient $-\phi$ is set to 0.5.\\

In the second step, the consideration set $A$ is leveraged to produce the Hessian matrix $M$. To do so, I implement the approach in Tian et al. (2021) to produce said matrix based on co-purchase regularities (approach discussed in greater detail in the Empirical Application section). The approach allows both $A$ and $M$ to be indirectly inferred from the same underlying simulated preferences. The inverse matrix itself is 50\% said specification, 50\% randomised, to reflect possible measurement error.\\

In the last step, I compute four different models, underpinned by two market structures: multi-good monopoly and single-good monopolistic competition. The models are: baseline, with outcomes derived from an unconstrained Slutsky matrix where the aggregate consideration set does not bind; unconstrained, where $A$ is less-than-full row rank; and lastly two setups with outcomes derived from the constrained Slutsky matrix - one where only non-negativity binds, and one where $A$ is also less-than-full row rank.\\

After each simulation draw, all models are compared across a range of characteristics. Given the price elasticity of demand in each draw, I determine own-price elasticity, cross-price elasticity, average magnitude of substitutes' and complements' cross-price elasticities, and the average number of complementary and substitute products per good, well as outcomes: average demand, consumer surplus, and aggregate profits.\\

\begin{table}[H] 
\centering
\footnotesize\singlespacing
\setlength{\tabcolsep}{6pt}
\renewcommand{\arraystretch}{1.2}
\label{tab:sim_consumer_choice}
\caption*{\textbf{Table}: Simulation of consumer choice counterfactuals}
\begin{tabular}{p{0.3\linewidth} r r c r}
\toprule
\textbf{Metric} & \textbf{Unconstrained} & \textbf{Baseline} & \textbf{Relation} & \textbf{\% draws} \\
\midrule
Avg own-price effect      &  $-0.37$  & $-0.53$  & $<$ (abs.)& 100   \\
Avg cross-price effect       & $0.008$  & $0.0186$ & $<$  & 100   \\
Avg substitution effect          & $0.0339$ & $0.0323$ & $>$  & 75.9  \\
Avg complementary effect  &  $-0.0502$& $-0.0199$& $>$ (abs.) & 100  \\
Avg number of substitutes      & $4.7$    & $5.2$    & $<$& 88.3  \\
Avg number of complements       & $2.1$   & $1.7$   & $>$& 88.3  \\
\midrule
\textbf{Metric} & \textbf{Const. (NN-LF)}& \textbf{Const. (NN)} & \textbf{Relation} & \textbf{\% draws} \\
\midrule
Avg own-price effect      &  $-0.37$  & $-0.53$  & $<$ (abs.)&  100  \\
Avg cross-price effect       & $0.008$  & $0.0186$ & $<$  &  100  \\
Avg substitution effect          & $0.0339$ & $0.0323$ & $>$  &  75.6 \\
Avg complementary effect  &  $-0.0502$& $-0.0199$& $>$ (abs.) & 99.9 \\
Avg number of substitutes      & $4.7$    & $5.2$    & $<$&  88 \\
Avg number of complements       & $2.1$   & $1.7$   & $>$&  88.2 \\
\midrule
\midrule
Firm consumer surplus          & $89.6$   & $178$  & $<$  & 100  \\
Monopoly consumer surplus      & $106.5$  & $107.3$  & $<$  & 94.8  \\
Firm aggregate profit           & $371.4$  & $391$  & $<$  & 56.6 \\
Monopoly aggregate profit       & $425.9$  & $429.3$  & $<$  & 100  \\
Firm aggregate demand           & $2.60$   & $3.73$   & $<$  & 100  \\
Monopoly aggregate demand       & $2.88$   & $2.90$   & $<$  & 93.7  \\
\bottomrule
\end{tabular}\\
\vspace{-0.5cm}\bigskip\RaggedRight\singlespacing
\footnotesize{Notes: "NN-LF" refers to the setting where both non-negativity and the vector space constraints imposed by the aggregate consideration set apply. "NN" refers to the setting where only non-negativity applies. "Relation" compares the outcomes of consumer choice under consideration set constraints to the unconstrained setting value. Rows marked "(abs.)" compare absolute values. "\% draws" is the percentage of simulation draws in which the stated relation (weakly) holds. No metric has an equality relation between models over 15\% of all draws.$^{***}p<0.01$, $^{**}p<0.05$, $^{*}p<0.1$.}
\end{table}

A few results stand out. Firstly, the constraints imposed on aggregate demand by a representative consumer's corner solutions do not appear to qualitatively affect outcomes.\footnote{This may nonetheless differ depending on the number and type of consumers which induce such corners.} Instead, a set of suggestive relationships between models bounded or not by whether $A$ is full row rank or not emerges.\\

Unsurprisingly, the absolute value of the average own-price effect is (weakly) lower when choice is constrained by the representative consumer's consideration set restriction on the consumption vector space. This is as predicted earlier, though slightly stricter.\\

More novel: the average cross-price effect is (weakly) lower when $A$ is less-than-full row rank. This constitutes one of the more elusive proofs of interest to us. Overall, the attenuation effect found in own-price effects appears to extends to cross-price effects.\\

The following offers some suggestions on the drivers of this result: the average magnitude of cross-price effects from complementary goods is (weakly) greater when choice is constrained by the representative consumer's consideration set restriction on the consumption vector space. Restricting consumption bundles to a consideration set drives an increase in complementarity, caused by regular co-consideration between goods. These appears to be another side of the attenuation effect - product demand is driven by demand of those goods it is co-considered with most often.\\

This is paired with a few rarer (but not so rare) occurrences. The average magnitude of cross-price effects from substitute goods is tendentially greater when $A$ is less-than-full row rank. On one hand, substitution effects grow stronger, much like complementarity. This is intuitive. Conditional on shared complements, the new competitive margin is in what to pair said complement with. In full-rank models, the decision to buy one good has a broader impact on those not bought; but in a setting where pairing one good with another more directly implies a decision not to buy yet another good that the first is often co-considered with strengthens the rivalry between the chosen and not-chosen good. On the other hand: the average number of complementary goods per product is generally greater when choice is constrained by the representative consumer's consideration set restriction on the consumption vector space.; and the average number of substitute goods per product is generally lower when $A$ is less-than-full row rank. In other words, on the net, the complementarity effect takes precedence in the cross-price outcome. \\

Amongst the market outcomes, note the following. Consumer surplus is (weakly) lower when choice is constrained by the representative consumer's consideration set restriction on the consumption vector space, in the single-product firm setting; but only tendencially lower in the same circumstances in the multi-product monopoly setting. Note also that product ownership concentration appears to lead to greater consumer surplus where the consideration set restriction on the consumption vector space binds in a majority of simulation draws, a result virtually absent in product choice. This suggests that the above-mentioned complementarities created by co-consideration are best taken advantage of in concentrated markets where incentives are internalised.\\

Aggregate profit is tendencially lower when $A$ is less-than-full row rank, in the multi-product monopoly setting. Surprisingly, however, under the same conditions, single-product firms yield greater aggregate profits in only a small majority of cases. This may be driven by the two possibilities discussed above: weaker own-price effects result in higher pricing power and thus higher prices; smaller cross-price effects - which as we saw are driven by a greater (lesser) number of complements (substitutes) and complementary (substitution) effects - diminish competitive concerns. The strategic complementary characteristic of Bertrand competition does the rest. Nonetheless, aggregate profits are always greater under multi-product monopoly than in single-product oligopolistic settings.\\

Lastly, average demand is tendencially lower when choice is $A$ is less-than-full row rank, across both single-product oligopolistic and multi-product monopoly settings. Monopoly aggregate demand is greater in the multi-product monopoly, in keeping with our discussion above.\\

The rational inattention literature consistently finds that behavioural restrictions to consumer choice reduce price elasticity (Matějka and MacKay, 2015; Caplin, Dean, and Leahy, 2019). This is intuitive for our setting too. Where choices are restricted the aggregate consideration set, for every consideration set alternative its next best alternative is on average worse than if there was no such restriction on the consumption bundles. Unlike in the existing literature, this effect is not necessarily driven by an "information gap", but can instead be a consequence of the underlying structure of preferences itself.\\

On the whole, when the $A$ is less-than-full row rank, own-price effects are weaker than they would otherwise be, as goods' demand becomes relatively more dependent on those goods with which they are often co-purchased. Often co-considered goods behave as complements, while goods often paired with the same products face substitution effects conditional on these. From these, consumer choice conditioned by a consideration set also translates into worse market outcomes for both consumers and firms than previous theory otherwise assumed; and it favours product ownership concentration at a greater clip, with the average draw repeatedly finding greater aggregate demand, consumer surplus, and aggregate profits in this setting.\\

I conclude with an illustrative example. At a base level, some goods effectively become more or less attractive, closer or farther substitutes, or even complementary, as cross-price effects become intertwined with often-co-purchased goods through considerations set constraints. Let

\begin{equation}
    A_{\substack{\{red\ onion,\ avocado,\\begin{equation}2pt]\ lime\}
             \times choices}}=
\begin{pmatrix}
2&0&1&2\\
2&2&2&4\\
0&2&1&2\\
\end{pmatrix}
\end{equation}

One approach for producing a substitute matrix consistent with the observed consideration set $A_{{\{red\ onion,\ avocado,\ lime\}
\times choices}}$ (Tian et al., 2021) produces the following matrices:

\begin{equation}
        M^{-1}=\begin{pmatrix}
        1.140 & -0.107 & -0.370\\
        -0.107 & 1.030 & -0.107\\
        -0.370 & -0.107 & 1.140
    \end{pmatrix}, \
    A(A'MA)^+A'=\begin{pmatrix}
        0.884 & 0.257 & -0.627\\
        0.257 & 0.514 & 0.257\\
        -0.627 & 0.257 & 0.884
    \end{pmatrix}
\end{equation}

Notice how, despite all three items behaving as substitutes in the first matrix, avocado becomes a complement to all in the second matrix. Red onions and lime, meanwhile, become stronger substitutes. This seems to be driven by avocado being in all consideration set alternatives. On what guacamole is concerned, avocado is more likely to be the one driving the purchases - sometimes with red onion, sometimes with lime, sometimes with both. All feasible ways for the consumer to adjust what they buy run through it. Once choice is constrained, this fact makes avocado a complement to both red onions and lime; conditional on it, the relevant margin is what to pair it with, milk or pasta. This residual rivalry explains the strengthening of substitute effects. The flipping of avocado from a substitute to a complement explains the rest. Effectively, regular co-consideration appears to drive complementarities, while strengthening substitution between goods which share co-considered products.\\

Following the example, we can also calculate aggregate profits, consumer surplus, and average price and demand across goods and market structures. At the Bertrand (Nash) equilibrium across both single-product firms and monopolist pricing:

\begin{table}[H]
\caption*{\textbf{Table}: Welfare outcomes under various market structure and consideration set constraint specifications.}
\centering
\begin{tabular}{lcc}
\toprule
$\Pi$/CS & \textbf{Consideration-set-adjusted} & \textbf{Baseline} \\
\midrule
$K$ firms & 20.22/0.78 & 20.36/1.57 \\
Monopolist & 20.54/1.03 & 21.41/1.07    \\
\bottomrule
\end{tabular}
\end{table}

\begin{table}[H]
\caption*{\textbf{Table}: Average market outcomes under various market structure and consideration set constraint specifications.}
\centering
\begin{tabular}{lcc}
\toprule
$\overline{\boldsymbol{p}^*}$/$\overline{\boldsymbol{q}^*}$ & \textbf{Consideration-set-adjusted} & \textbf{Baseline} \\
\midrule
$K$ firms & 11.269/0.598 & 7.865/0.863
  \\
Monopolist & 9.997/0.685 & 10/0.714  \\
\bottomrule
\end{tabular}
\end{table}

All of these match what we observed in the Monte Carlo simulation.
\pagebreak

\subsection{Welfare analysis}

Representation rationalisation is weaker than a full aggregation result, with implications for calculating social welfare. To see this, note:

\begin{equation}
\begin{aligned}
\sum_{i=1}^N U_i(\boldsymbol{Q}/N,W)
&= \sum_{i=1}^N [(\boldsymbol{Q}/N)'(\boldsymbol{\delta}_i+\phi_i \boldsymbol{p}) - \frac{1}{2}(\boldsymbol{Q}/N)' M (\boldsymbol{Q}/N)]+\sum_{i=1}^N\phi_iW \\
&= (\boldsymbol{Q}/N)'(\boldsymbol{\delta}+\phi \boldsymbol{p}) - \sum_{i=1}^N \frac{1}{2}(\boldsymbol{Q}/N)' M (\boldsymbol{Q}/N)+\phi W \\
&= (\boldsymbol{Q}/N)'(\boldsymbol{\delta}+\phi \boldsymbol{p}) - N \cdot \frac{1}{2}(\boldsymbol{Q}/N)' M (\boldsymbol{Q}/N)+\phi W  \\
&= (\boldsymbol{Q}/N)'(\boldsymbol{\delta}+\phi\boldsymbol{p}) - \frac{1}{2N} \boldsymbol{Q}' M \boldsymbol{Q}+\phi W \\
& = \frac{1}{N}U_{RC}(\boldsymbol{Q},W)
\end{aligned}
\end{equation}

for $W=\sum_{i=1}^Nw_i$. Therefore:

\begin{equation}
    U_{RC}(\boldsymbol{Q},W)\propto\sum_{i=1}^N U_i(\boldsymbol{Q}/N,W)\neq \sum_{i=1}^NU_i(\boldsymbol{q}_i)
\end{equation}

with proportionality also a result of utility functions being unique only up to a scale. Defining $U_{RC}$ over the aggregates $\boldsymbol{\delta}$ and $\phi$ does not deliver aggregation at the utility level.\\


Up to an affine transformation, it can nonetheless be shown that $U_{RC}(\boldsymbol{Q})$ is proportional to the utilitarian planner’s maximum attainable  social welfare $SW$ as a function of $\boldsymbol{Q}$:

\begin{equation}
\begin{aligned}
SW(\boldsymbol{Q})
&= \max_{\{\boldsymbol{q}_i\}_{i=1}^N:\sum_{i=1}^N \boldsymbol{q}_i = \boldsymbol{Q}}\sum_{i=1}^N(\boldsymbol{q}_i'\boldsymbol{\delta}_i-\frac12 \boldsymbol{q}_i' M \boldsymbol{q}_i) \\
&\Leftrightarrow
\boldsymbol{q}_i^*(\boldsymbol{Q})=(\boldsymbol{Q}/N)+M^{-1}(\boldsymbol{\delta}_i-\boldsymbol{\bar\delta}),\ \forall i
\end{aligned}
\end{equation}

Then:

\begin{equation}
\begin{aligned}
SW^*(\boldsymbol{Q})
&= \sum_{i=1}^N(\boldsymbol{q}_i^{*}(\boldsymbol{Q})'\boldsymbol{\delta}_i-\frac12\, \boldsymbol{q}_i^*(\boldsymbol{Q})' M \boldsymbol{q}_i^*(\boldsymbol{Q})) \\
&= \boldsymbol{Q}'\boldsymbol{\bar\delta} - \frac{1}{2N}\boldsymbol{Q}' M \boldsymbol{Q}
\;+\;\frac12\sum_{i=1}^N(\boldsymbol{\delta}_i-\boldsymbol{\bar\delta})' M^{-1}(\boldsymbol{\delta}_i-\boldsymbol{\bar\delta}) \\
&= \sum_{i=1}^N U_i(\boldsymbol{Q}/N)
+\frac12\sum_{i=1}^N(\boldsymbol{\delta}_i-\boldsymbol{\bar\delta})' M^{-1}(\delta_i-\boldsymbol{\bar\delta})\\
&= N \cdot U_{RC}(\boldsymbol{Q})
+[\text{constant in }\boldsymbol{Q}\ge 0]
\end{aligned}
\end{equation}

In other words, the utility obtained by a representative consumer with aggregate parameters $\boldsymbol{\delta}$ and $\phi$ can be used to study how changes in aggregate demand impact the social welfare obtained by a social planner finding the best achievable welfare via transfers. This makes the framework well-suited for social welfare comparisons across policies that operate primarily through changes in aggregate consumption $\boldsymbol{Q}$.\\

\subsection{Details on usage of transaction sample data}

Because, by \textbf{Proposition 2} and the Minkowski sum, $\boldsymbol{z}$ is unique in $\mathrm{col}(A)$, $T$ is also unique, for $T=\sum_{i=1}^NT_i$. However, because only $\boldsymbol{Q}$ is identified and not $\boldsymbol{Q}/T$, $T$ is not identified. Therefore, in practice, the only implication of $T$ for an analysis of aggregated demand is that, should we want to estimate parameters for the average demand per transaction yet only observe a random sample of transactions $T_r$, we can divide the resulting $\boldsymbol{Q}_r$ by $T_r$ to obtain:

\begin{equation}
    \frac{1}{T_r}\boldsymbol{Q}_r=\frac{1}{T_r}(\frac{T_r}{T}\boldsymbol{Q})=\frac{1}{T}\boldsymbol{Q}
\end{equation}

This is our setting in this paper, such that all estimated parameters must be understood as relative to the average transaction.\\

\pagebreak

\subsection{Accounting for multi-product monopolistic competition}

Two extreme market structures we have considered earlier in this Appendix are those of a finite number of single-product firms, and of a single multi-product monopoly. I now extend the analysis to incorporate multi-product monopolistic competition, whereby simultaneous multi-product and single-product firms compete with each other. More specifically, I provide a condition under which a unique equilibrium can be computed for these markets. Discussion for the model's implications for entry dynamics is left for future work.\\

For a more general settings with multiple single- and multi-product firms, I propose imposing a bound on the extent to which cross-price effects from goods owned by the same multi-product firm are accounted for by firms:

\begin{equation}
W=\Omega+M^{-1}+\alpha(M^{-1}\circ G)
\end{equation}
\\
Then, if $M^{-1}_{ii}+\Omega_{ii}=2M_{ii}^{-1}>\sum_{j\neq i:G_{ij}=0}|M^{-1}_{ij}|+(1+\alpha)\sum_{j\neq i:G_{ij}=1}|M^{-1}_{ij}|, \ \forall i$, or:


\begin{equation}
    \alpha<\min_i\frac{2M_{ii}^{-1}-\sum_{j\neq i:G_{ij}=0}|M^{-1}_{ij}|}{\sum_{j\neq i:G_{ij}=1}|M^{-1}_{ij}|}-1
\end{equation}

then the resulting demand Jacobian implies a well-behaved demand system.\footnote{This is effectively a strict diagonal dominance condition.} Alternatively, we have a both sufficient and necessary condition in:


\begin{equation}
    \alpha<\min_{\boldsymbol{x}\neq 0,\boldsymbol{x}'(M^{-1}\circ G)\boldsymbol{x}<0}\frac{\boldsymbol{x}'(\Omega+M^{-1})\boldsymbol{x}}{-\boldsymbol{x}'(M^{-1}\circ G)\boldsymbol{x}}
\end{equation}

This approach is less conservative but may be harder to calculate in some settings.\\

It may be of interest to readers to know that Federgruen and Hu (2021) already solves the problem of demand invertibility for a broad class of symmetric, binary ownership matrices by effectively permuting the product ordering so that $G+I$ becomes block diagonal. Block diagonal matrices are invertible if each of their individual diagonal blocks is invertible. Conveniently, this is always the case for symmetric, binary ownership matrices. What I propose above is a general solution to the problem under any ownership matrix, likely pertinent for settings such as e.g. common ownership. 

\pagebreak

\subsection{Retail stock-outs}

Expressions derived from the consumer choice model in Chapter 1 can be particularly useful to understand the impact on retail settings where a given product goes out of stock. We return to the matter of price equilibria again at the end of the next section. \\

\textit{\textbf{Lemma A.1}: Following the stock-out of a good sold by a monopolist retailer, the resulting changes in optimal demand, aggregate profits, and consumer surplus have the following closed-form expression:}

\begin{equation}
    \Delta \boldsymbol{Q}^*=-\frac{1}{4}\begin{pmatrix}  \sigma_{ii}\delta_i+\boldsymbol{\sigma}_i'\boldsymbol{\delta}_{-i}\\ \boldsymbol{\sigma}_i\delta_i\end{pmatrix}
\end{equation}

\begin{equation}
    \Delta \Pi =- \frac{1}{4\phi}\big[\sigma_{ii}\delta_i^2+2\delta_i(\boldsymbol{\sigma}_i'\boldsymbol{\delta}_{-i})\big]
\end{equation}

\begin{equation}
    \Delta CS = -\frac{3}{8}\big[\sigma_{ii}\delta_i^2+2\delta_i(\boldsymbol{\sigma}_i'\boldsymbol{\delta}_{-i})\big]
\end{equation}

\textit{for $\Sigma = A(A'MA)^+A'$.}\\

\textit{Proof}: All model outputs are in one of the following forms: $\boldsymbol{y_1}=\alpha\boldsymbol{x}'\Sigma\boldsymbol{x}$ or $\boldsymbol{y_2}=\alpha\Sigma\boldsymbol{x}$ for $\alpha$ a strictly positive scalar, $\boldsymbol{x}$ a strictly positive vector, and $\Sigma$ a positive semi-definite matrix. Suppose we wish to consider the impacts derived from the removal of the $i$-th component of $\boldsymbol{x}$ and $\Sigma$. These terms can be expressed as such:

\begin{equation}
    \boldsymbol{x}= \begin{pmatrix}  x_i\\ \boldsymbol{x}_{-i}\end{pmatrix}
\end{equation}

and

\begin{equation}
    \Sigma= \begin{pmatrix}  \sigma_{ii} & \boldsymbol{\sigma}_i'\\ \boldsymbol{\sigma}_i & \Sigma_{-i}\end{pmatrix}
\end{equation}

The notation for $\boldsymbol{\sigma}_i$ excludes the element corresponding to $\sigma_{ii}$. Starting with $\boldsymbol{y_1}$, we may restate it as:

\begin{equation}
    \boldsymbol{y_1} = \sigma_{ii}x_i^2+2x_i(\boldsymbol{\sigma}_i'\boldsymbol{x}_{-i})+\boldsymbol{x}_{-i}\Sigma{-i}\boldsymbol{x}_{-i}
\end{equation}

of which the first two terms are the contribution from the $i$-th dimension of both $\boldsymbol{x}$ and $\Sigma$. If we remove the $i$-th product, i.e. the $i$-th dimension of $\boldsymbol{x}$ and $\Sigma$, the difference between the original and reduced term is:

\begin{equation}
    \Delta_1 =-\alpha\big[\sigma_{ii}x_i^2+2x_i(\boldsymbol{\sigma}_i'\boldsymbol{x}_{-i})\big]
\end{equation}

We can similarly re-write $\boldsymbol{y_2}$:

\begin{equation}
    \boldsymbol{y_2}=\alpha\begin{pmatrix}  \sigma_{ii}x_i+\boldsymbol{\sigma}_i'\boldsymbol{x}_{-i}\\ \boldsymbol{\sigma}_ix_i+\Sigma_{-i}\boldsymbol{x}_{-i}\end{pmatrix}
\end{equation}

If the $i$-th component is dropped, the first (upper) block drops out, and the lower block's reduced version is $\alpha\Sigma_{-i}\boldsymbol{x}_{-i}$. \qedwhite\\


Even where $M$ is symmetric and diagonally dominant (common assumptions, sufficient for positive definiteness), consumer surplus and aggregate profits may or may not \textit{decrease} as a result of product exit, and the ways in which they do so will depend on the structure of the representative consumer's consideration set. The QQUM specification aids us in allowing the expressions above to be written in closed form.  This may not be the whole story however; a product whose removal raises aggregate profit may nonetheless be kept within the assortment if it can minimise the impact of other products' removal. As can be seen above, the removal of any given product is exacerbated or dampened by its interaction with others. A correct understanding of the interaction matrix is important for addressing stock-outs.\\

\pagebreak

\subsection{Empirical estimation for less-than-full row $A$}

We have assumed thus far that, for an appropriate proxy $W$, we have that $M^{-1}\approx\sum_{j=0}^J\alpha_jW^j$. What is less clear is whether $W$ is an appropriate proxy for $M^{-1}$ or if it should be used to approximate $A(A'MA)^+A'$ as a whole where $A$ is less-than-full row rank. This is a matter better left to a test: model both cases and verify which fits the data best. It is less so a problem in the latter case, as our estimation approach above already fits that alternative without any additional changes. The former, however, requires expressing $A(A'MA)^+A'$ as a function of $M^{-1}$, so that $\sum_{j=0}^J\alpha_jW^j$ can be fit. We could use $M=I-\alpha W$ but the pseudo-inversion step would not be able to follow the Neumann series, introducing a bottleneck on estimation.\\

To achieve this, note:\\

\textit{\textbf{Lemma}:  $A(A'MA)^+A'=M^{-1}A(A'M^{-1}A)^+A'M^{-1}$, for $M$ a positive definite matrix.}\hfill $\blacksquare$\\

\textit{Proof}: The intermediate step is to show that $A(A'MA)^+A'M=M^{-1}A(A'M^{-1}A)^+A'$; if this is the case, both sides of the equation can be multiplied by $M^{-1}$ to yield the Lemma's equality. I do so by noting that both sides of the equation are an $M$-orthogonal projection onto the same space $col(A)$. Since orthogonal projections are unique, the two must be equivalent. An orthogonal projection to space $col(A)$ must satisfy three conditions: the matrix must be in $col(A)$, be idempotent, and self-adjoint in $M$. \\

For a vector $\boldsymbol{u}=A\boldsymbol{v}\in col(A)$, $A(A'MA)^+A'M(Av)=Av$. Similarly, noting that $col(M^{-1}A)=col(A)$ as $M^{-1}$ is positive definite with $col(M^{-1})=\mathbb{R}$, we have $\boldsymbol{u}=M^{-1}A\boldsymbol{v}\in col(A)$ such that $M^{-1}A(A'M^{-1}A)^+A'(M^{-1}Av)=M^{-1}Av$. This means that every vector in $col(A)$ is also in the column space of these expressions. The proof is complete by noting the reverse -  the column space of these expressions is also a subset of $col(A)$: $A(A'MA)^+A'M=AB$ for $B=(A'MA)^+A'M$, and $col(AB)\subseteq col(A)$; $M^{-1}A(A'M^{-1}A)^+A'=M^{-1}AB$ for $B=(A'M^{-1}A)^+A'$, with $col(M^{-1}AB)\subseteq col(M^{-1}A)=col(A)$. Idempotency is similarly straightforward: $A(A'MA)^+A'M\cdot A(A'MA)^+A'M=A(A'MA)^+A'M$ and $M^{-1}A(A'M^{-1}A)^+A'\cdot M^{-1}A(A'M^{-1}A)^+A' = M^{-1}A(A'M^{-1}A)^+A'$. Lastly, note:

\begin{equation}
    (A(A'MA)^+A'M\boldsymbol{x})'M\boldsymbol{y}=\boldsymbol{x}'M(A(A'MA)^+A'M\boldsymbol{y})
\end{equation}

and

\begin{equation}
    (M^{-1}A(A'M^{-1}A)^+A'\boldsymbol{x})'M\boldsymbol{y}=\boldsymbol{x}'M(M^{-1}A(A'M^{-1}A)^+A'\boldsymbol{y})
\end{equation} \qedwhite

If $A(A'MA)^+A'=M^{-1}A(A'M^{-1}A)^+A'M^{-1}$, then we can implement grid search to jointly estimate the parameters in $M^{-1}\approx\sum_{j=0}^J\alpha_jW^j$: set initial values for $\alpha_j,\forall j$, calculate $\hat{M^{-1}}A(A'\hat{M^{-1}}A)^+A'\hat{M^{-1}}$, and regress the model to obtain $\boldsymbol{\hat{\beta}}$ and $\hat{\phi}$. The optimal $\boldsymbol{\alpha}$ will be those which best fits the data.\\

\pagebreak

\subsection{Participation rates}

In our empirical application, all stores were at least 1km away from the nearest store competitor at the beginning of the sample in 2020, except for Portimão  (700m away). Maia faced entry by a competitor 700m away from its store in February 2023, just after the end of our sample, and another 1.4km away in late February 2022; Mafra faced (re-)entry 300m away in February 2021; and Portimão faced an additional entry 600m away in June 2022. This high level of activity in the industry reflects a recent expansion strategy pursued by European discount retailers in the Portuguese market, which remained active throughout the COVID pandemic. Given this thesis's focus on intra-supermarket competition, these may raise concerns that inter-supermarket competition impacts our outcomes. \\

In addition, in the standardised regressions above, all product quantities purchased are corrected by the number of transactions observed each period in the relevant supermarket. This is in line with linear demand aggregation. However, such correction introduces endogeneity concerns: if prices at the grocery store increase, consumers may decide to shop elsewhere, and vice versa. What may be the implications for mark-ups? Start from the differentiated-product, supermarket-time specific, Bertrand-Nash, Lerner index:

\begin{equation}
    L_{m,t} = - ((G \circ E_{m,t})^{-1})\cdot \boldsymbol{1}
\end{equation}

where $G$ is the ownership matrix, and 

\begin{equation}
    E_{m,t} = J_{m,t} \cdot p_{m,t}/s_{m,t}
\end{equation}  

with $s$ the per-customer demand, and $J=\frac{\partial s}{\partial p}$, each specified per market $m$ and time period $t$. If participation $N$ varies with prices, the per-customer Jacobian we estimated earlier from demand, $\hat J_{m,t} = \frac{\partial s}{\partial p}$,
bundles two margins:

\begin{equation}
    \frac{\partial s}{\partial p}=(1/N)\cdot\frac{\partial q}{\partial p}-s \cdot(\psi^N)'
\end{equation}

where $\psi^N_j \equiv \frac{\partial \ln N}{\partial p_j}$ (participation sensitivity). Thus the intensive-margin Jacobian needed for mark-ups is:

\begin{equation}
    J^{int}_{m,t} = \hat J_{m,t} + s_{m,t} \cdot (\psi^N_{m,t})'
\end{equation}

If $\frac{\partial q}{\partial p}<0$ (higher prices reduce participation), ignoring participation would make own-price slopes look too small in magnitude, which would inflate the Lerner indices. Adding $s\cdot(\psi^N)'$ would be necessary to restore the intensive margin, typically lowering the Lerner index relative to the uncorrected approach.\\

In this section, I discuss the matter of participation rates and verify the robustness of our estimates of price elasticity of demand both to inter-supermarket competition and price endogeneity.

\subsubsection{Transaction trends}

Consider the time trends for the number of transactions observed per supermarket in the sample period:

\begin{figure}[H]
    \centering
    \caption*{\textbf{Figure}: Per-supermarket transaction trends}
    \includegraphics[scale=0.6]{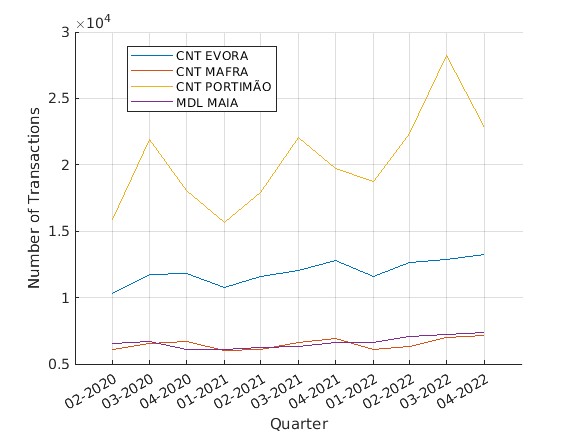}
\end{figure}

The trends do not resemble those observed for revenue-weighted mean prices in the paper. Nonetheless, the lack of a clear relationship between the data on grocery stores' prices and transactions numbers may mask a participation elasticity. Core inflation rose substantially in Portugal through the period of analysis. Nominal inflation pushed all prices up, but if as a result people shifted spending from elsewhere to groceries, this may have led to more transactions in-store. For example, the price of home food relative to restaurant meals may have fallen, attracting shoppers despite higher nominal grocery prices. This could be masking the correct price elasticities of grocery store purchases.

\subsubsection{Estimation of participation rate elasticity}

We wish to identify the causal effect of grocery store prices on participation $N$, net of common inflation
and fixed market traits. The proposed estimation strategy is a first-difference IV panel estimator:

\begin{equation}
    \Delta\ln(N_{m,t})=\delta \cdot \Delta \ln P_{m,t}  + \tau_t + \Delta u_{m,t}
\end{equation}

Parameter $\delta$ is the participation elasticity. The number of transactions is calculated on a per-quarter, per-supermarket basis, while the price index $P_{m,t}$ is set under two specifications: (i) a simple average across goods; and (ii) a revenue-weighted function of per-quarter, per-supermarket goods prices. This approach matches our earlier empirical specification, which used quarterly data, but also limits the size of the regression we can run. Variable $\tau_t$ is a time trend.\\

Taking first differences removes latent time-invariant terms while preempting trend drifts. An instrument is necessary in this setting, as before, to isolate the price effect on participation rates with exogenous variation, avoiding reverse causality. I use the change in a good's competitors' average price (our previous instrument), aggregated into an index as that which it instruments for. \\

For robustness, I consider current, 1-2 lags and select that with the highest first-stage F. With one endogenous regressor, we run 2SLS and report Eicker-Huber-White standard errors; the 2SLS p-value; the Anderson-Rubin p-value, which stays valid even when instruments are weak; and the (homoskedastic) first-stage F for the chosen instrument set. Results are shown below:\\

\begin{table}[H]\centering
\label{tab:iv_results}
\caption*{\textbf{Table}: Participation shares - IV results by specification}
\begin{threeparttable}
\begin{tabular}{lccccc}
\toprule
Price index & $\hat\delta$ (SE) & $p$-value & AR $p$-value & First-stage $F$ & $N$ \\
\midrule
Simple average            & 0.056\ (2.685)  & 0.983 & 0.981 & 6.920  & 32 \\
Revenue-weighted   & $-2.855$\ (4.890) & 0.559 & 0.223 & 6.119  & 32 \\
\bottomrule
\end{tabular}
\end{threeparttable}
\vspace{-0.5cm}\medskip\\\bigskip\RaggedRight\singlespacing
\footnotesize Notes: $\hat\delta$ is the IV estimate; standard errors in parentheses. "AR $p$-value" is the Anderson-Rubin test $p$-value. "First-stage $F$" is the first-stage F-statistic. $N$ is the number of observations.
\end{table}

The preferred instrument is the first lag of the competitor price change with a first-stage $F\approx6.12$ and $6.92$ respectively, still however indicating weak instrument strength. Nonetheless, neither specification shows a detectable effect of prices on participation, even under the Anderson-Rubin test. The simple-average specification yields $\hat\delta\approx0.06$ and the revenue-weighted specification $\hat\delta\approx-2.86$. Deflating the price indices by the CPI leaves the results unchanged.\\

Though underpowered, the results suggest participation sensitivity is not a significant driver of our results. This seems to confirm that (i) inter-supermarket competition is not a major problem in the set of stores and period I have sampled, and (ii) adjusting demand by the number of transactions observed each period in each store raises no endogeneity concerns within our sample. Nonetheless, the approach used here could be averaging out what are likely important effects amongst staple goods in any supermarket assortment. The reader is therefore encouraged to perceive this thesis's mark-up estimates as upper bounds, especially so for goods on the upper tail of the revenue distribution.\\

\end{document}